\documentclass[preprint,showpacs,preprintnumbers,amsmath,amssymb,natbib]{revtex4}

\usepackage{graphicx}% Include figure files
\usepackage{epsfig}		
\usepackage{dcolumn}% Align table columns on decimal point
\usepackage{bm}% bold math
\usepackage{subfigure}
\def\0{\mbox{\tiny $0$}}
\def\1{\mbox{\tiny $1$}}
\def\2{\mbox{\tiny $2$}}
\def\3{\mbox{\tiny $3$}}
\def\4{\mbox{\tiny $4$}}
\def\5{\mbox{\tiny $5$}}
\def\6{\mbox{\tiny $6$}}
\def\7{\mbox{\tiny $7$}}
\def\8{\mbox{\tiny $8$}}
\def\9{\mbox{\tiny $9$}}

\def\f14{\mbox{\tiny $\frac{1}{4}$}}

\begin{document}

\title{Probing phase-space noncommutativity through quantum beating, missing information and the thermodynamic limit}

\author{A. E. Bernardini}
\email{alexeb@ufscar.br}
\altaffiliation[On leave of absence from]{~Departamento de F\'{\i}sica, Universidade Federal de S\~ao Carlos, PO Box 676, 13565-905, S\~ao Carlos, SP, Brasil.}
%\altaffiliation{Also at Instituto de F\'{\i}sica Gleb Wataghin, UNICAMP, PO Box 6165, 13083-970, Campinas, SP, Brasil}
\author{O. Bertolami}
\email{orfeu.bertolami@fc.up.pt}
\altaffiliation[Also at~]{Instituto de Plasmas e Fus\~ao Nuclear, Instituto Superior T\'ecnico, Av. Rovisco Pais, 1, 1049-001, Lisboa.} 
\affiliation{Departamento de F\'isica e Astronomia, Faculdade de Ci\^{e}ncias da
Universidade do Porto, Rua do Campo Alegre 687, 4169-007, Porto, Portugal.}
\date{\today}% It is always \today, today,

\begin{abstract}
In this work we examine the effect of phase-space noncommutativity on some typically quantum properties such as quantum beating, quantum information, and decoherence.
To exemplify these issues we consider the two-dimensional noncommutative quantum harmonic oscillator whose components behavior we monitor in time.
This procedure allows us to determine how the noncommutative parameters are related to the missing information quantified by the linear quantum entropy and by the mutual information between the relevant Hilbert space coordinates.
Particular questions concerning the thermodynamic limit of some relevant properties are also discussed in order to evidence the effects of noncommutativity.
Finally, through an analogy with the Zeeman effect, we identify how some aspects of the axial symmetry of the problem suggest the possibility of decoupling the noncommutative quantum perturbations from unperturbed commutative well-known solutions.
\end{abstract}

\date{\today}
\maketitle

\section{Introduction}

Loss of quantum coherence, quantum entanglement, quantum beating, wave function collapse and several related issues are at the heart of quantum features that must be necessarily considered in order to construct a suitable quantum measurement theory \cite{01A,02A,03A}.
These effects allow for a better understanding of the frontiers between quantum and classical descriptions \cite{02A} as well as, for instance, on the generation of superposition states of mesoscopic systems, known as Schr\"odinger {\em cat states} \cite{04A,05A}.

The central question examined in this work concerns typically quantum mechanical features in the context of a phase-space noncommutative (NC) extension of quantum mechanics (QM) \cite{Catarina,OB01,Catarina001,Catarina002}.
More particularly, the effect of phase-space noncommutativity on phenomena such as quantum beating, missing information and decoherence.

To start with, noncommutativity in the configuration space has been firstly suggested as a way to regularize quantum field theory \cite{Snyder47}.
Furthermore, it is a long-held belief that in quantum theories which include gravity, the nature of the space-time is modified at the Planck scale, and that noncommutativity must be considered at high energy scales.
Noncommutative geometry is also encountered in the context of string theory/M-theory \cite{Connes,Douglas,Seiberg}.
We also mention that several aspects of NC theories have been studied, including noncommutative gauge and field theories \cite{Nekrasov01,Konechny00,Connes94,GraciaBondia,Szabo,Martinetti}.

In the quantum mechanical context, NC extensions have been discussed with respect to the quantum Hall effect \cite{Prange,Belissard}, for electrons in a magnetic field projected to the lowest level in the Landau problem in the phase-space \cite{Nekrasov01}, for the $2$-dimensional (2D) quantum harmonic oscillator in Cartesian coordinates \cite{Rosenbaum}, and in the gravitational quantum well for ultra-cold neutrons \cite{06A,07A}.
The NC QM is supported by canonical extensions of the Heisenberg-Weyl algebra \cite{06A,07A,08A,09A,Gamboa,Guisado}.
The theory lives in a $2d$-dimensional phase-space where time is required to be a commutative parameter, and coordinate and momentum variables obey a NC algebra (c. f. Eq.~(\ref{Eq31})).

It is important to stress that phase-space NC extensions of QM have striking features.
These include putative violations of the Robertson-Schr\"odinger uncertainty relation \cite{Catarina001}, a feature that has somehow been observed in a recent experiment \cite{Stein}.
Moreover, the regularizing features of phase-space NC QM are remarkable and shown in minisuperspace quantum cosmology models \cite{Bastos00}, and in black-hole physics where a $L^2$ wave-function can be found and the singularity problem can be resolved \cite{Bastos004,Bastos005}.
 
The phase-space NC extension of QM is more suitably formulated in the Weyl-Wigner-Groenewold-Moyal (WWGM) formalism for QM \cite{Groenewold,Moyal,Wigner}.
Its framework inspired on the understanding of the QM statistical structure has been developed in the last years \cite{Catarina,Dunne,Jellal,Horvathy1,Ho,Bolonek,Chaichian,Nair,Djemai,Dias1,Pool,Dubin,Wilde}.
It is thus fairly natural to use the WWGM formalism in our study of the 2D quantum harmonic oscillator.
As will be seen, NC introduces several conceptual subtleties in describing the dynamics of quantum systems and affect some typical quantum phenomena such as quantum beating, ensued missing information and wave function distortions.
The quantum beating related to the interchange of information between the Hilbert sub-spaces of the problem are, as will be seen, exclusively due to the NC properties.
Not less important, is the missing information due to probability distribution distortions, which can be displayed through the coupling of NC degrees of freedom to a thermal reservoir.

The outline of this manuscript is as follows.
In section II, we review the Weyl-Wigner formalism and its applicability in describing the NC extension of the QM in the phase-space.
We report about the formal procedure for obtaining the energy spectrum and the Wigner {\em stargen}functions of a NC quantum system through the Groenewold-Moyal {\em star}-product in the NC phase-space that exhibit both coordinate and momentum noncommutativity.
In section III, we apply the formalism to calculate the NC properties of the quantum harmonic oscillator on the plane.
Besides reproducing some results on the {\em star}genvalue problem of the harmonic oscillator in the NC phase-space, our emphasis is on obtaining the phase-space time-evolution that leads to the time-dependence of the Wigner function.
This provides a fruitful framework for investigating NC effects.
We show how noncommutativity potentially affects the quantum coherence of a state vector that is obtained through an arbitrary quantum preparation procedure.
Hints about the possibility of observing noncommutativity through quantum beating effects are discussed.
The missing information quantified by the linear quantum entropy is discussed in section IV.
The quantum entropy destroyed by the NC character is quantified and a well-defined prescription for the wave function distortion in thermodynamic equilibrium is obtained.
Results for the thermodynamic limit of the internal energy, Boltzmann entropy, and heat capacity are derived.
The quantum issue of the 2D NC harmonic oscillator is reformulated in section V.
It is shown that some aspects of the axial symmetry of the problem allow for a factorization of the NC effects.
Obtaining a well-defined quantum number and a factorizable wave-function component for noncommutativity are shown to be relevant for quantifying the NC effects on the 2D quantum harmonic oscillator.
Finally, we draw our conclusions in section VI and address some related issues concerning the anisotropic NC harmonic oscillator, and possible extra linear contributions to the potential in space coordinates.

\section{The Weyl transform and the Wigner function in the phase-space}

The predictive power of QM relies on its formalism to connect quantum observables to expectation values.
In this respect, the Wigner function and the Weyl transform establish an alternative formulation to the better known Schr\"odinger and Heisenberg frameworks.
This formalism provides an interesting insight into the nature of QM and its classical limit.
The Wigner representation has also the operational advantage of exhibiting all the information about the state vector in the phase-space. 
This contrasts with the conventional QM approach which provides information about position, or momentum, but not both together.
This formulation is akin to the formalism of statistical mechanics.

The inception of the Wigner function is the definition of the Weyl transform of an operator $\hat{O}(\hat{q}, \hat{p})$ through the expression
\begin{equation}
O^W(q, p) =\hspace{-.2cm} \int \hspace{-.15cm}dy\,\exp{\left[i \,p\, y/\hbar\right]}\,\langle q - y/2 | \hat{O} | q + y/2 \rangle = \int \hspace{-.15cm} dy \,\exp{\left[- i \,q\, y/\hbar\right]}\,\langle p - y/2 | \hat{O} | p + y/2\rangle,
\end{equation}
in both coordinate and momentum basis, where operators, $\hat{q}$ and $\hat{p}$, are converted into $c$-numbers, $q$ and $p$.
The fundamental property of the Weyl transform is that the trace of the product of two operators, $\hat{O}_1$ and $\hat{O}_2$, is given by the phase-space integral of the product of their Weyl tranforms \cite{Wigner,Case} as 
\begin{equation}
Tr_{\{q,p\}}\left[\hat{O}_1\hat{O}_2\right] = h^{-1} 
\int \hspace{-.15cm}\int \hspace{-.15cm} dq\,dp \,O^W_1(q, p)\,O^W_2(q, p),
\end{equation}
where $h$ is Planck's constant.
The trace properties can be related to the density matrix properties of the state vector described by $\hat{\rho}$: 
\begin{equation}
Tr_{\{q,p\}}\left[\hat{\rho}\hat{O}\right] =  \langle O \rangle = h^{-1} 
\int \hspace{-.15cm}\int \hspace{-.15cm} dq\,dp \,\rho^W(q, p)\,{O^W}(q, p).
\end{equation}
By identifying the density matrix $\hat{\rho}$ with $|\Psi \rangle \langle \Psi |$, one can define the Wigner function,
\begin{equation}
W(q, p) =  h^{-1} \rho^W = 
\int \hspace{-.15cm}dy\,\exp{\left[i \, p \,y/\hbar\right]}\,
\Psi(q - y/2)\,\Psi^{\ast}(q + y/2),
\end{equation}
which can be naturally generalized to a statistical mixture, such that the expectation value of an observable $\hat{O}$ can be computed through 
\begin{equation}
\langle O \rangle = 
\int \hspace{-.15cm}\int \hspace{-.15cm} dq\,dp \,W(q, p)\,{O^W}(q, p).
\end{equation}
By identifying $\Psi(q)$ with
\begin{equation}
\Psi(q) =  h^{-1}\int \hspace{-.15cm} dp\,\exp{\left[i \, p \,q/\hbar\right]}\,\Phi(p),
\end{equation}
one straightforwardly obtains probability distributions for $q$ and $p$ as
\begin{equation}
\int \hspace{-.15cm} dq\,W(q, p) = \Phi^{\ast}(p)\,\Phi(p)
\quad\mbox{and}\quad
\int \hspace{-.15cm} dp \,W(q, p)= \Psi(q)^{\ast}\Psi(q),
\end{equation}
i. e. the Wigner function represents the distribution in the phase-space represented by $\Psi(q)$, and can also be computed from $\Phi(p)$ as
\begin{equation}
W(q, p) = 
\int \hspace{-.15cm}dy\,\exp{\left[-i\, q \,y/\hbar\right]}\,
\Phi(p - y/2)\Phi^{\ast}(p + y/2).
\end{equation}
Additional properties related to the density matrix theory can be obtained from the above prescription \cite{Hillery,Lee}.
Our interest is mainly concerned with the definition of $Tr[\hat{\rho}^2]$ as
\begin{equation}
Tr[\hat{\rho}^2] =  h^{-1}\int \hspace{-.15cm}\int \hspace{-.15cm} dq\,dp \,W(q, p)^2,
\end{equation}
which demands for an extra normalization factor, $h^{-1}$, in order to keep consistency with the density matrix theory that sets $Tr[\hat{\rho}^2] = Tr[\hat{\rho}] = 1$ for pure states.

Providing a distribution in the phase-space and a representation for the expectation values of quantum observables is not, nevertheless, an exclusive property of the Wigner functions.
A similar procedure could be implemented through other function candidates to provide enough information about the quantum systems.
Given that it admits positive and negative amplitude values, the Wigner function cannot be strictly interpreted as a probability distribution.
On its hand, the Husimi distribution, for instance, is an alternative which ensures non-negative values and admits a probability interpretation \cite{Husimi}.
This distribution is largely used in quantum optics \cite{Carmichael} and to study quantum effects in superconductors \cite{Callaway}.
Likewise, to describe light in optical phase-space, a third proposal, the Sudarshan-Glauber $P$ representation \cite{Glauber} is a common tool.

Despite these examples, due to its simpler formulation, the Weyl-Wigner formalism is the most often procedure used to overcome some obstacles in quantum dynamics.
To provide an example of its functionality, let us recover the commutative Heisenberg-Weyl algebra of the ordinary QM,
\begin{equation}
\left[ \hat{Q}_i,  \hat{Q}_j \right] = 0 , \hspace{0.5 cm} \left[ \hat{Q}_i,  \hat{P}_j \right]
= i \hbar \delta_{ij} , \hspace{0.5 cm} \left[ \hat{P}_i,  \hat{P}_j \right] = 0 ,
\hspace{0.5 cm} i,j= 1, ... ,d ~. 
\label{NCeq}
\end{equation}
The Weyl transform can be extended to a $d$-dimensional phase-space, $\{\mathbf{Q},\mathbf{P}\}$, such that any quantum operator, $O(\mathbf{Q},\mathbf{P}; t)$, in terms of $c$-numbers, can be written as
\begin{equation}
O^W(\mathbf{Q},\mathbf{P}; t) = \int\hspace{-.1cm}\int\hspace{-.1cm}d\mathbf{x}\,d\mathbf{y}
F(\mathbf{x},\mathbf{y}; t) \, \exp{\left[\frac{i}{\hbar}\left(\mathbf{x}\cdot\mathbf{P} +\mathbf{y}\cdot\mathbf{Q}\right)\right]},
\label{WigWig}
\end{equation}
where, through the $d$-dimensional generalization of the Weyl transform, one has
\begin{equation}
F(\mathbf{x},\mathbf{y}; t) = h^{-d}\, Tr_{\{\mathbf{Q},\mathbf{P}\}}
\left[O(\mathbf{Q},\mathbf{P}; t)\,\exp{\left[\frac{i}{\hbar}\left(\mathbf{x}\cdot\mathbf{P} +\mathbf{y}\cdot\mathbf{Q}\right)\right]}
\right].
\label{WeyWey}
\end{equation}
The algebra in Eq.~(\ref{NCeq}) is recovered through the introduction of the Groenewold-Moyal {\em star}-product defined on the space of commutative functions as
\begin{equation}
h \star  g = \exp{\left[\frac{\Lambda}{2}\epsilon_{ij}\partial_{r_i}\partial_{s_j}\right]}h(r)\,g(s)|_{r=s},
\end{equation}
where $\Lambda$ is a suitable constant.

The main feature of WWGM phase-space formalism is that
\begin{equation}
\langle \Psi |O_{1} ( {\mathbf Q}, {\mathbf P}; t) O_{2}
( {\mathbf Q}, {\mathbf P}; t) | \Psi \rangle 
= \int\hspace{-.15cm} \int \hspace{-.15cm} 
d{\mathbf P}\, d{\mathbf Q} \,
\rho^{W}(\mathbf{Q},\mathbf{P}; t)\,
O^{W}_{1} (\mathbf{Q},\mathbf{P}; t) \star  O^{W}_{2}(\mathbf{Q},\mathbf{P}; t), 
\label{wigner}
\end{equation}
where 
\begin{equation}
\rho^{W}(\mathbf{Q},\mathbf{P}; t) =h^{-d}
\int d{\mathbf z} \exp[\frac{i}{\hbar} {\bf z}\cdot{\mathbf P}]\langle
{\mathbf Q} -\frac{{\mathbf z}}{2} |\rho |{\mathbf Q} + \frac{{\mathbf
z}}{2}\rangle, 
\label{dist}
\end{equation}
which shares the same properties of the Wigner quasi-probability distribution function from Eq.~(\ref{WigWig}), once the density matrix for a pure quantum state, $\rho =|\Psi\rangle\langle \Psi|$ is introduced.
The Moyal bidifferential {\em star}-operator is given by 
\begin{equation}
\star =  \exp \left[
\frac{i\hbar}{2} ( {\overleftarrow{\nabla}_{{\mathbf Q}}}\cdot
{\overrightarrow{\nabla}_{{\mathbf P}}} - {\overleftarrow{\nabla}_{{\mathbf
P}}}\cdot {\overrightarrow{\nabla}_{{\mathbf Q}}} ) \right],
\label{moyal} 
\end{equation}
such that the dynamics of the Heisenberg operator is denoted by
\begin{equation}
\dot{O}^W( {\mathbf Q}, {\mathbf P}; t) = -\frac{i}{\hbar} \left[O^W(0), H^W( {\mathbf Q}, {\mathbf P}; t)\right]_{\star} = \frac{i}{\hbar} (H^W \star O^W - O^W\star H^W),
\label{moyal2} 
\end{equation}
which follows naturally from the WWGM formalism.
From the above construction, the commutative variables ${\mathbf P}$ and ${\mathbf Q}$ satisfy the Hamilton equations of motion, namely
\begin{eqnarray}
\dot{\mathbf P} &=& -\frac{i}{\hbar} \left[p^W(0), H^W( {\mathbf Q}, {\mathbf P}; t)\right]_{\star} = - \vec{\nabla}_{\mathbf Q} H^W; \nonumber\\
\dot{\mathbf Q} &=& -\frac{i}{\hbar} \left[Q^W(0), H^W( {\mathbf Q}, {\mathbf P}; t)\right]_{\star} = + \vec{\nabla}_{\mathbf P} H^W,
\label{cnumber} 
\end{eqnarray}
and, therefore, the corresponding classical dynamics interpretation is inherent from the formalism.

Once again, the Wigner function projection on configuration and momentum spaces yields the quantum mechanical configuration and momentum probabilities, respectively.
Although somewhat beyond the scope of this work, several additional properties can be recovered from the WWGM formalism \cite{Wigner,Catarina,Rosenbaum}.
For instance, by making use of the integral representation, one has
\begin{equation}
\begin{split}
    O^{W}_{1} ( {\mathbf Q}, {\mathbf P}) \star O^{W}_{2} ( {\mathbf Q}, {\mathbf P})  =
    (2\pi\hbar)^{-2d}\int\dots\int d{\mathbf P}^{\prime} \ d{\mathbf P}^{\prime\prime}
      d{\mathbf Q}^{\prime} \ d{\mathbf Q}^{\prime\prime}  O^{W}_{1} ({\mathbf P}^{\prime}, {\mathbf Q}^{\prime})
       O^{W}_{2} ({\mathbf P}^{\prime\prime}, {\mathbf Q}^{\prime\prime})\\
       \exp[-\frac{2i}{\hbar}({\mathbf P}\cdot({\mathbf Q}^{\prime} - {\mathbf Q}^{\prime\prime})
       + {\mathbf P}^{\prime}\cdot({\mathbf Q}^{\prime\prime} - {\mathbf Q}) +
        {\mathbf P}^{\prime\prime}\cdot({\mathbf Q} - {\mathbf Q}^{\prime}))], \label{int1}\\
\end{split}
\end{equation}
and obtains that
\begin{equation}
\int\int d{\mathbf P} \ d{\mathbf Q} \ H^{W} ( {\mathbf Q}, {\mathbf P}) \star\rho^{W} ( {\mathbf Q}, {\mathbf P})
= \int\int d{\mathbf P} \ d{\mathbf Q} \ H^{W} ( {\mathbf Q}, {\mathbf P})
 \rho^{W} ( {\mathbf Q}, {\mathbf P}) = E, \label{int2}
\end{equation}
which shall be useful in the following {\em star}genvalue problems involving NC variables.

Turning to the extended NC algebra, the following commutation relations are satisfied
\begin{equation}
\left[ \hat{q}_i,  \hat{q}_j \right] = i \theta_{ij} , \hspace{0.5 cm} \left[ \hat{q}_i,  \hat{p}_j \right] = i \hbar \delta_{ij} ,
\hspace{0.5 cm} \left[ \hat{p}_i,  \hat{p}_j \right] = i \eta_{ij} ,  \hspace{0.5 cm} i,j= 1, ... ,d
\label{Eq31}
\end{equation}
where $\eta_{ij}$ and $\theta_{ij}$ are invertible antisymmetric real constant ($d \times d$) matrices, one can define the matrix
\begin{equation}
\Sigma_{ij} \equiv \delta_{ij} + {1\over\hbar^2}  \theta_{ik} \eta_{kj},
\label{Eq32}
\end{equation}
which is  an equally invertible if $\theta_{ik}\eta_{kj} \neq -\hbar^2 \delta_{ij}$.
Under a linear transformation this algebra can be mapped into the usual Heisenberg-Weyl algebra as in Eq.~(\ref{NCeq}) via the Seiberg-Witten (SW) map \cite{Seiberg}, which can be cast in the form
\begin{equation}
 \hat{q}_i = A_{ij}  \mathit{\hat{Q}}_j + B_{ij}  \hat{\Pi}_j \hspace{1 cm}
 \hat{p}_i = C_{ij}  \mathit{\hat{Q}}_j + D_{ij}  \hat{\Pi}_j,
\label{Eq35}
\end{equation}
where ${\bf A},{\bf B},{\bf C},{\bf D}$ are real constant matrices. 
The above transformation constrained by the NC relations from Eq.~(\ref{NCeq}) is easily shown to obey the following matrix equations \cite{Rosenbaum}
\begin{equation}
{\bf A} {\bf D}^T - {\bf B} {\bf C}^T = {\bf I}_{d \times d} \hspace{1 cm} {\bf A} {\bf B}^T - {\bf B} {\bf A}^T = {1\over\hbar} {\bf \Theta} \hspace{1 cm}
{\bf C} {\bf D}^T - {\bf D} {\bf C}^T = {1\over\hbar} {\bf N}~,
\label{Eq36}
\end{equation}
where ${\bf A}, {\bf B}, {\bf C}, {\bf D}, {\bf \Theta}, {\bf N}$ are matrices with entries $A_{ij}, B_{ij}, C_{ij}, D_{ij} , \theta_{ij} , \eta_{ij}$ and the superscript $T$ stands for matrix transposition, respectively. 

The linear transformation ensures that the abovementioned NC algebra admits a representation in terms of the Hilbert space of ordinary QM.
In the noncommutativity context, the Wigner-function is much more than a visual insight into the meaning of quantum states.
As it will be pointed out in the following section, the Weyl-Wigner formalism is actually the most natural approach for implementing the NC formalism.

\section{The time-dependence of the 2D NC harmonic oscillator: quantum beating and decoherence}

The generalization of the Weyl functions and the Wigner distributions to the NC phase-space
and the general features of the Weyl-Wigner map, which includes its covariant generalization to $d$-dimensional systems, have been formally developed in Ref.~\cite{Catarina}. 
The generalized {\em star}-product, the extended Moyal bracket and the NC Wigner function have been mathematically formulated, and it is found that the extended Weyl-Wigner map is independent of any particular choice of the SW map.

It is not our purpose to re-discuss here the mathematical grounds of the formalism and its general applicability to quantum systems, as performed in \cite{Catarina}, but in order to relate the noncommutativity properties with observable quantum phenomena, it will suffice to apply its results to the well-known harmonic oscillator on the plane.

The quantum mechanical problem of the 2D harmonic oscillator is the most accessible system for which the NC phase-space properties can be probed \cite{Catarina,Rosenbaum}, and quantum effects quantified.

Thus, let us consider the quantum Hamiltonian,
\begin{equation}
\hat{H}_{HO}(\mathbf{q},\mathbf{p}) = \frac{\mathbf{p}^2}{2m} + \frac{1}{2}m \omega^2 \mathbf{q}^2,
\end{equation}
on the NC $x-y$ plane, with spatial and momentum noncommutativity being set by
\begin{equation}
\left[  \hat{q}_i,  \hat{q}_j \right] = i \theta \epsilon_{ij}\hspace{0.2 cm}, \hspace{0.2 cm} \left[  \hat{q}_i,  \hat{p}_j \right] = i \delta_{ij}\hbar 
%\left(1 - \frac{\theta\eta}{4\hbar^2}\right)\hspace{0.2 cm}
, \hspace{0.2 cm} \left[  \hat{p}_i,  \hat{p}_j \right] = i \eta \epsilon_{ij}\hspace{0.2 cm}, \hspace{0.2 cm} i,j=1,2
\end{equation}
with $\epsilon_{ij} = -\epsilon_{ij}$, such that the map to the commutative operators is given by
\begin{eqnarray}
\mathit{\hat{Q}}_i &=& \mu \left(1 - {\theta \eta\over\hbar^2} \right)^{- 1 / 2} \left( \hat{q}_i + {\theta\over 2 \lambda \mu \hbar} \epsilon_{ij}  \hat{p}_j \right)~,\nonumber\\
\hat{\Pi}_i &=& \lambda \left(1 - {\theta \eta\over\hbar^2} \right)^{-1 / 2} \left( \hat{p}_i-{\eta\over 2 \lambda \mu \hbar} \epsilon_{ij}  \hat{q}_j \right),
\label{SWinverse}
\end{eqnarray}
through the SW map,
\begin{equation}
 \hat{q}_i = \lambda  \mathit{\hat{Q}}_i - {\theta\over2 \lambda \hbar} \epsilon_{ij} {\Pi}_j \hspace{0.5 cm},\hspace{0.5 cm}  \hat{p}_i = \mu {\Pi}_i + {\eta\over 2 \mu \hbar} \epsilon_{ij}  \mathit{\hat{Q}}_j~,
\label{SWmap}
\end{equation}
which is invertible when the parameters $\lambda$ and $\mu$ are constrained by the relationship
\begin{equation}
{\theta \eta \over 4 \hbar^2} = \lambda \mu ( 1 - \lambda \mu ),
\label{constraint}
\end{equation}
with $\theta\eta \lesssim \hbar^2$, and with the corresponding Jacobian reading
\begin{equation}
{\partial (q,p)\over \partial (\mathit{Q}, \Pi)} = (\det {\mathbf \Omega})^{1/2}=1 - {\theta \eta\over\hbar^2}.
\end{equation}

The Hamiltonian in terms of the commutative variables, $\mathit{\hat{Q}}_i$ and $\hat{\Pi}_i$, reads: 
\begin{equation}
H^W_{HO}(\mbox{\bf \em Q},\mathbf{\Pi}) = \alpha^2\mbox{\bf \em Q}^2 +\beta^2\mathbf{\Pi}^2 + \gamma \sum_{i,j = 1}^2{\epsilon_{ij}\hat{\Pi}_i \mathit{\hat{Q}}_j},
\label{Hamilton}
\end{equation}
where 
\begin{eqnarray}
{\alpha}^2 &\equiv& {\lambda^2 m \omega^2\over 2} + {\eta^2\over 8m \mu^2 \hbar^2}~,\nonumber\\
{\beta}^2 &\equiv& {\mu^2\over 2m} + {m \omega^2 \theta^2\over 8 \lambda^2 \hbar^2}~,\nonumber\\
{\gamma} &\equiv& \frac{\theta}{2\hbar}m \omega^2  + \frac{\eta}{2m\hbar}.
\label{param}
\end{eqnarray}

Giving that, from Eq.~(\ref{cnumber}), the commutative variables, $\mbox{\bf \em Q}$ and $\mathbf{\Pi}$, satisfy the Hamilton equations of motion, one obtains the following set of coupled first-order differential equations,
\begin{eqnarray}
\dot{\Pi}_k &=& -\frac{i}{\hbar} \left[\Pi_k,\,H^W_{HO}\right] = -2 \alpha^2\,\mathit{Q}_k - \gamma\,\varepsilon_{jk}\Pi_j,\nonumber\\
\dot{\mathit{Q}}_k &=& -\frac{i}{\hbar} \left[\mathit{Q}_k,\,H^W_{HO}\right] =  ~~2 \beta^2\,\Pi_k - \gamma\,\varepsilon_{jk}\mathit{Q}_j,\qquad k,j = 1,2
\label{eqs01}
\end{eqnarray}
so that $\mbox{\bf \em Q}$ and $\mathbf{\Pi}$ may be interpreted as classical dynamical variables within the WWGM formalism.
The above equations can be rewritten as two uncoupled forth-order differential equations as
\begin{eqnarray}
\ddddot{\Pi}_k &=&  -2 (\gamma^2 + 4\alpha^2\beta^2)\,\ddot{\Pi}_k + (\gamma^2 - 4\alpha^2\beta^2)\,\Pi_k, \nonumber\\
\ddddot{\mathit{Q}}_k &=&  -2 (\gamma^2 + 4\alpha^2\beta^2)\,\ddot{\mathit{Q}}_k + (\gamma^2 - 4\alpha^2\beta^2)\,\mathit{Q}_k,
\label{eqs01}
\end{eqnarray}
from which one gets the solutions
\small
\begin{eqnarray}
\mathit{Q}_1(t)&=& x\,\cos(\Omega t)\cos(\gamma t) + y\,\cos(\Omega t)\sin(\gamma t)
  +\frac{\beta}{\alpha}\left[\pi_y\,\sin(\Omega t)\sin(\gamma t) + \pi_x\,\sin(\Omega t)\cos(\gamma t)
\right],\nonumber\\
\mathit{Q}_2(t)&=& y\,\cos(\Omega t)\cos(\gamma t) - x\,\cos(\Omega t)\sin(\gamma t)
 -\frac{\beta}{\alpha}\left[\pi_x\,\sin(\Omega t)\sin(\gamma t) - \pi_y\,\sin(\Omega t)\cos(\gamma t)
\right],\nonumber\\
\Pi_1(t)&=& \pi_x\,\cos(\Omega t)\cos(\gamma t) + \pi_y\,\cos(\Omega t)\sin(\gamma t)
 -\frac{\alpha}{\beta}\left[y\,\sin(\Omega t)\sin(\gamma t) + x\,\sin(\Omega t)\cos(\gamma t)
\right],~~\nonumber\\
\Pi_2(t)&=& \pi_y\,\cos(\Omega t)\cos(\gamma t) - \pi_x\,\cos(\Omega t)\sin(\gamma t)
 +\frac{\alpha}{\beta}\left[x\,\sin(\Omega t)\sin(\gamma t) - y\,
\sin(\Omega t)\cos(\gamma t)
\right],~~
\label{solutions}
\end{eqnarray}
\normalsize
where $x,\, y,\,\pi_x,$ and $\pi_y$ are arbitrary parameters, and 
\begin{equation}
\Omega   = 2 \alpha \beta  =  \omega \sqrt{(2\lambda\mu - 1)^2 + \varepsilon^2},
\label{eq37}
\end{equation}
with
\begin{equation}
\varepsilon = \frac{1}{2\hbar}\left[m\omega\theta + \frac{\eta}{m\omega}\right],
\end{equation}
so that $\gamma = \omega \varepsilon$.
By conveniently noticing that $\Omega \sim \omega[1 + \mathcal{O}(\varepsilon^2)]\times\vert2\lambda\mu - 1\vert \sim \omega[1 + \mathcal{O}(\theta^2,\, \eta^2,\, \theta\eta)]$, one sees that the NC parameters, $\theta$ and $\eta$, introduce second-order modifications onto $\Omega$. 
Likewise, the modifications due to $\gamma = \omega \varepsilon$ correspond typically to first order effects.
Notice that by setting $\varepsilon = 0$ one recovers the solutions for the 2D harmonic oscillator with uncoupled $x-y$ coordinates.
The above results lead to time-invariant quantities expressed by,
\begin{eqnarray}
\sum_{i=1}^2{\left(\frac{\alpha}{\beta} \mathit{Q}_i(t)^2 + \frac{\beta}{\alpha}\Pi_i(t)^2\right)} &=& \frac{\alpha}{\beta} (x^2 + y^2) + \frac{\beta}{\alpha}(\pi_x^2 + \pi_y^2),\nonumber\\
\sum_{i,j=1}^2{\left(\epsilon_{ij} \mathit{Q}_i(t)\,\Pi_j(t)\right)} &=& x\,\pi_y - y \,\pi_x, 
\label{invariant}
\end{eqnarray}
which will be useful in the subsequent discussion.

\begin{figure}
\vspace{-3.0 cm}
\includegraphics[width= 8cm]{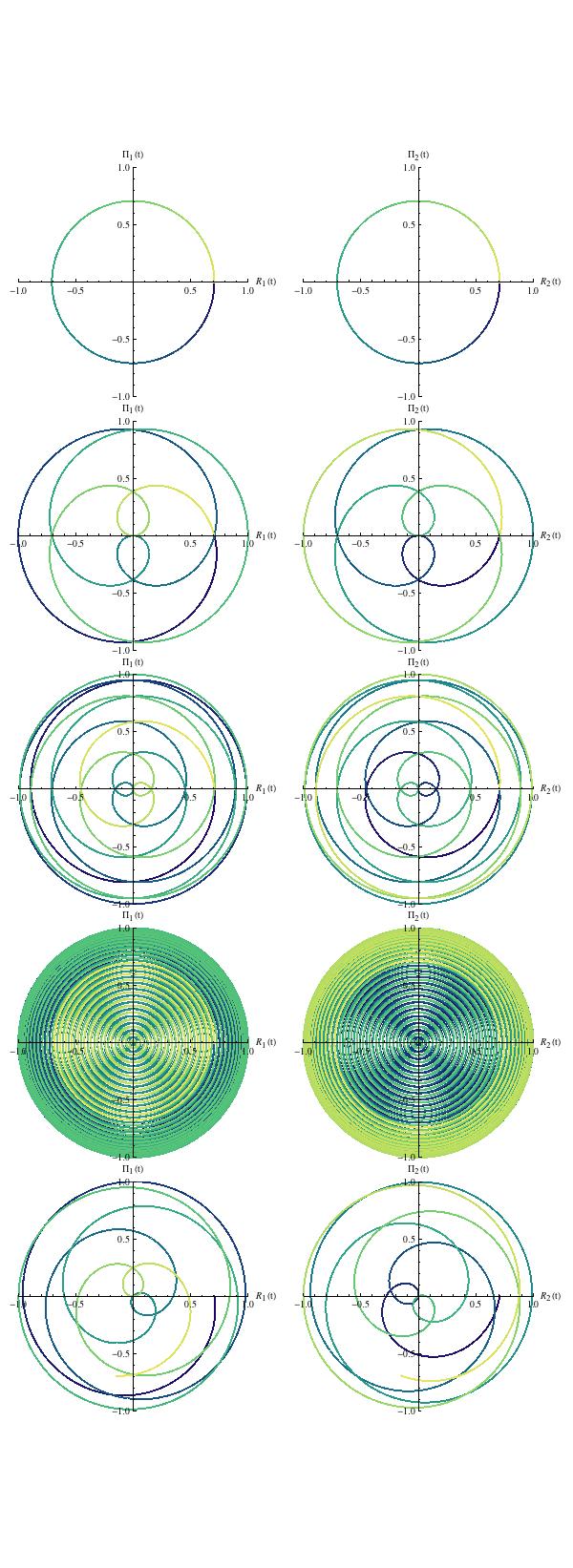}
\includegraphics[width= 1.5cm]{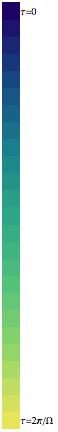}
\caption{\footnotesize  (Color online) Time evolution of the phase-space coordinates,$(\mathit{Q}_1(t),\Pi_1(t))$ and $(\mathit{Q}_2(t),\Pi_2(t))$, for the NC harmonic oscillator. 
The first plot line refers to the phase-space elliptical orbits of commutative harmonic oscillators (as if one had set $\epsilon = 0$ in the NC map).
From the second to the forth plot lines one has set arbitrary values for $\epsilon$, $\epsilon = 1/4,\, 1/10$, and $1/100$ respectively. 
The {\em precession} motion describes close orbits for these cases.
One has used a {\em BlueGreenYellow} ({\em GrayLevel}) scale in order to denote the time scale, $\tau$, varying from $0$ (blue (dark gray)) to $2 \pi / \Omega$ (yellow (light gray)), such that orbits start and finish at $(x,\pi_x) = (y,\pi_y) = (1/\sqrt{2},0)$.
The last pair of plots corresponds to $\epsilon = 1/2\pi$, which does not result into a closed orbit since it is not a rational number.
By convenience, one has set $\alpha = \beta$, that is equivalent to $m\omega = \hbar = 1$ }
\label{Phase}
\end{figure}

The time evolution of the phase-space coordinates,$(\mathit{Q}_1(t),\Pi_1(t))$ and $(\mathit{Q}_2(t),\Pi_2(t))$, for the NC harmonic oscillator are depicted in Fig.~\ref{Phase}.
We have considered the time scale in the range $[0, 2\pi/\Omega]$.
For convenience, we define an auxiliary variable, $\epsilon$,
\begin{equation}
\epsilon = {\gamma\over\Omega} = \frac{\varepsilon}{\sqrt{(2\lambda\mu - 1)^2 + \varepsilon^2}},
\end{equation}
in order to perform a non-perturbative analysis of the results.
If $\epsilon$ can be written as a rational number, one encounters a beating effect along the phase-space trajectories. 
The same does not occur for non-rational values of $\epsilon$ (c. f. the last plot in Fig.~\ref{Phase}).

Following the notation of Section II, the {\em stargen}functions for the Hamiltonian problem from Eq.~(\ref{Hamilton}) are obtained from the {\em stargen}value equation,
\begin{equation}
H^W_{HO} \star \rho_{n_{\tiny 1},n_{\tiny 2}}^W (\mbox{\bf \em Q},\mathbf{\Pi}) = 
E_{n_{\tiny{1}},n_{\tiny{2}}}\,\rho^W_{n_{\tiny{1}},n_{\tiny{2}}} (\mbox{\bf \em Q},\mathbf{\Pi}),
\label{help01}
\end{equation}
which, from the analysis developed in Ref.~\cite{Rosenbaum}, results into
\begin{equation}
\rho_{n_{\tiny{1}},n_{\tiny{2}}}^W (\mbox{\bf \em Q},\mathbf{\Pi}) =  \frac{(-1)^{n_1+n_2}}{\pi^2\hbar^{2}}\exp\left[{-\frac{1}{\hbar}\left(\frac{\alpha}{\beta}\mbox{\bf \em Q}^2 + \frac{\beta}{\alpha}{\mathbf{\Pi}}^2\right)}\right] \, L^0_{n_1} \left(\Omega_{+}/\hbar\right) \,L^0_{n_2}\left(\Omega_{-}/\hbar\right),
\label{Lague01}
\end{equation}
where $L^0_n$ are the associated Laguerre polynomials, $n_1$ and $n_2$ are non negative integers, and
\begin{equation}
{\Omega}_{\pm} = {\alpha\over\beta}\mbox{\bf \em Q}^2 + {\beta\over\alpha}\mathbf{\Pi}^2 \mp 2 \sum_{i,j = 1}^2{\left(\epsilon_{ij}\Pi_i \mathit{Q}_j\right)},
\end{equation}
such that the energy spectrum is given by
\begin{equation}
E_{n_{\tiny 1},n_{\tiny 2}} = \hbar\left[2\alpha\beta(n_1 + n_2 + 1) + \gamma (n_1 - n_2)\right],
\end{equation}
and one has
\begin{equation}
\int^{^{+\infty}}_{_{-\infty}}\hspace{-.5cm}d\mathit{Q}_1\int^{^{+\infty}}_{_{-\infty}}\hspace{-.5cm} d\Pi_1\int^{^{+\infty}}_{_{-\infty}}\hspace{-.5cm} d\mathit{Q}_2\int^{^{+\infty}}_{_{-\infty}}\hspace{-.5cm} d\Pi_2 \,\,\rho_{n_{\tiny{1}},n_{\tiny{2}}}^W (\mbox{\bf \em Q},\mathbf{\Pi}) = 1.
\end{equation}

From the results of Ref.~\cite{Catarina}, one realizes, using the inverse SW map Eq.~(\ref{SWinverse}), the constraint Eq.~(\ref{constraint}), and the normalization $(\det {\bf \Omega})^{-1/2} = \left(1- \theta \eta / \hbar^2 \right)^{-1}$, that the reported NC Wigner function is independent of $\lambda, \mu$.
The implications of the above results for the wave function time evolution are discussed in the following.

\subsection{Quantum beating and decoherence}

The dynamical evolution of a state vector described by a generic Wigner function, $\rho^W (\mbox{\bf \em Q},\mathbf{\Pi})$, follows the motion of coordinates in the phase-space, $\{\mathit{Q}_i,\Pi_i\}$.
One has only to ensure that each point of the Wigner function moves in the correlated paths, $1 \leftrightarrow 2$, as depicted for instance in Fig.~\ref{Phase} for the NC harmonic oscillator.
This is reflected by a characteristic invariance property of Wigner functions that sets
\begin{equation}
\rho^W (\mbox{\bf \em Q},\mathbf{\Pi}) \equiv \rho^W (\mbox{\bf \em Q},\mathbf{\Pi}; t) =\rho^W (\{x,\pi_x\},\{y,\pi_y\};0),
\label{help02}
\end{equation}
as one can obtain from the inverse transformation derived from Eq.~(\ref{solutions}) \cite{Wigner}.
If the state vector preparation follows the prescription of the quantum numbers $n_1$ and $n_2$, from Eq.~(\ref{invariant}), one can verify that the condition prescribed by Eq.~(\ref{help02}) results into stationary Wigner functions, $\rho_{n_1,n_2}^W (\mbox{\bf \em Q},\mathbf{\Pi})$. 

However, the properties of the Wigner function are not constrained, in general, by the dynamics of the system.
For the NC harmonic oscillator, an analysis using the properties of the Laguerre polynomials (c. f. Eq.~(\ref{Lague01})) shows that some features are associated with the preparation of states rather than to the physics of its time evolution.

One should notice that, for instance, by rewriting the quantum numbers for the {\em star}genvalues through the constraint $n = n_1 + n_2$, the state vector can be prepared as a normalized statistical mixture with a well-defined quantum number, $n$, as
\begin{eqnarray}
\rho^W_n(\mbox{\bf \em Q},\mathbf{\Pi}) &=&\frac{(-1)^n}{1+n} \sum_{n_1=0}^n \rho^W_{n_{\tiny{1}},n_{\tiny{2}}}(\mbox{\bf \em Q},\mathbf{\Pi})\nonumber\\
         &=& \frac{1}{\pi^2\hbar^2} \frac{(-1)^n}{1+n}\exp\left[-\frac{1}{\hbar}\left(\frac{\alpha}{\beta}\mbox{\bf \em Q}^2 + \frac{\beta}{\alpha}{\mathbf{\Pi}}^2\right)\right] \,\sum_{n_1=0}^n L^0_{n_1} \left(\Omega_{+}/\hbar\right) \,L^0_{n - n_1}\left(\Omega_{-}/\hbar\right),
\label{help03}
\end{eqnarray}
from which, by noticing that
\begin{equation}
\sum_{n_1=0}^n L^c_{n_1} \left(a\right) L^d_{n - n_1}\left(b\right) = L^{c+d+1}_{n}\left(a + b\right),
\label{help04}
\end{equation}
one obtains
\begin{equation}
\rho^W_n (\mbox{\bf \em Q},\mathbf{\Pi}) \equiv \rho^W_n (\xi^2; 0) = \frac{1}{\pi^2\hbar^2} \frac{(-1)^n}{1+n} \exp{\left(-\xi^2/\hbar\right)}\,L^1_ n(2\xi^2/\hbar),
\label{help05}
\end{equation}
where
\begin{equation}
\xi^2 = \frac{\alpha}{\beta}\mbox{\bf \em Q}^2 + \frac{\beta}{\alpha}{\mathbf{\Pi}}^2 = \frac{\alpha}{\beta}(x^2 + y^2) + \frac{\beta}{\alpha}(\pi_x^2 + \pi_y^2).
\label{help05F}
\end{equation}
Results from section V suggest that such a normalized quantum superposition, $\rho^W_n (\mbox{\bf \em Q},\mathbf{\Pi})$, can be related to the solution of the commutative 2D harmonic oscillator written in terms of axially symmetric coordinates.
By using the same Laguerre polynomial identity from Eq.~(\ref{help04}), one can find the usual decomposition of 
$\rho^W_n (\mbox{\bf \em Q},\mathbf{\Pi})$ into orthogonal Cartesian coordinates, $x$ and $y$, such that
\begin{equation}
\rho^W_n (\mbox{\bf \em Q},\mathbf{\Pi}) \equiv \rho^W_n (\xi^2) = \frac{(-1)^n}{1+n} \sum_{n_x=0}^n 
\rho^W_{n_{\tiny{x}}}(\xi^2_x) \,\rho^W_{n - n_{\tiny{x}}}(\xi^2_y),
\label{help05B}
\end{equation}
represents the quantum superposition of normalized uncoupled pairs of one-dimensional (1D) harmonic oscillators, where
\begin{equation}
\rho^W_{n_{\tiny{x,y}}}(\xi^2_{x,y}) = \frac{1}{\pi\hbar} \exp{\left[-\xi^2_{x,y}/\hbar\right]}
L^0_{n_{x},n_{y}} \left(2\xi^2_{x,y}/\hbar\right),
\label{help05C}
\end{equation}
with $n_y = n - n_x$, and 
\begin{eqnarray}
\xi_{x}^2 &=& \frac{\alpha}{\beta}x^2 + \frac{\beta}{\alpha}\pi_x^2, \nonumber\\
\xi_{y}^2 &=& \frac{\alpha}{\beta}y^2 + \frac{\beta}{\alpha}\pi_y^2,
\end{eqnarray}
where we have used Eqs.~(\ref{help02}) and (\ref{help05F}).

Since the state vector described by Eq.~(\ref{help05}) does not exhibit any NC effect from Hamiltonian Eq.~(\ref{Hamilton}), then a quantum system can be arranged in order to exhibit the properties from the state described by Eq.~(\ref{help05}).
However, even for the NC phase-space, $\rho^W_n (\xi^2)$ describes a stationary state vector (c. f. Eq.~(\ref{invariant})).
The same is not true for the composition of state vectors defined by,
\begin{equation}
\rho^W_{n_{\tiny{x}},n_{\tiny{y}}}(\mbox{\bf \em Q},\mathbf{\Pi};t) = \rho^W_{n_{\tiny{x}}}(\xi^2_x) \,\rho^W_{n_{\tiny{y}}}(\xi^2_y) \rightarrow \rho^W_{n_{\tiny{x}}}(\mbox{\bf \em Q},\mathbf{\Pi};t) \,\rho^W_{n_{\tiny{y}}}(\mbox{\bf \em Q},\mathbf{\Pi};t).
\label{help05D}
\end{equation}
For the NC scenario described through the dynamics arising from Eq.~(\ref{solutions}) constrained by the invariance conditions, Eqs.~(\ref{help02}) and (\ref{help05F}), one can obtain the explicit time-dependence for the state vector from Eq.~(\ref{help05D}) by noticing that the time evolution of $\rho^W_{n_{\tiny{x,y}}}(\xi^2_{x,y})$ mix the components from $x$ and $y$ phase-spaces since one has that
\small
\begin{eqnarray}
\xi^2_{x} &\sim&\left[\left(\frac{\alpha}{\beta}\mathit{Q}_1^2 + \frac{\beta}{\alpha}\Pi_1^2\right)\cos{(\gamma t)}^2 + \left(\frac{\alpha}{\beta}\mathit{Q}_2^2 + \frac{\beta}{\alpha}\Pi_2^2\right)\sin{(\gamma t)}^2 - \left(\frac{\alpha}{\beta}\mathit{Q}_1 \mathit{Q}_2 + \frac{\beta}{\alpha}\Pi_1\Pi_2\right)\sin{(2\gamma t)}\right],~~~~~\nonumber\\
\xi^2_{y} &\sim&\left[\left(\frac{\alpha}{\beta}\mathit{Q}_1^2 + \frac{\beta}{\alpha}\Pi_1^2\right)\sin{(\gamma t)}^2 + \left(\frac{\alpha}{\beta}\mathit{Q}_2^2 + \frac{\beta}{\alpha}\Pi_2^2\right)\cos{(\gamma t)}^2 + \left(\frac{\alpha}{\beta}\mathit{Q}_1 \mathit{Q}_2 + \frac{\beta}{\alpha}\Pi_1\Pi_2\right)\sin{(2\gamma t)}\right].~~~~~
\label{help05G}
\end{eqnarray}
\normalsize
The above defined state vectors are stationary for the commutative case ($\gamma  = 0$), and they have indeed a time dependence mediated by $\gamma$.

\begin{figure}
\vspace{-1.2 cm}
\includegraphics[width= 10cm]{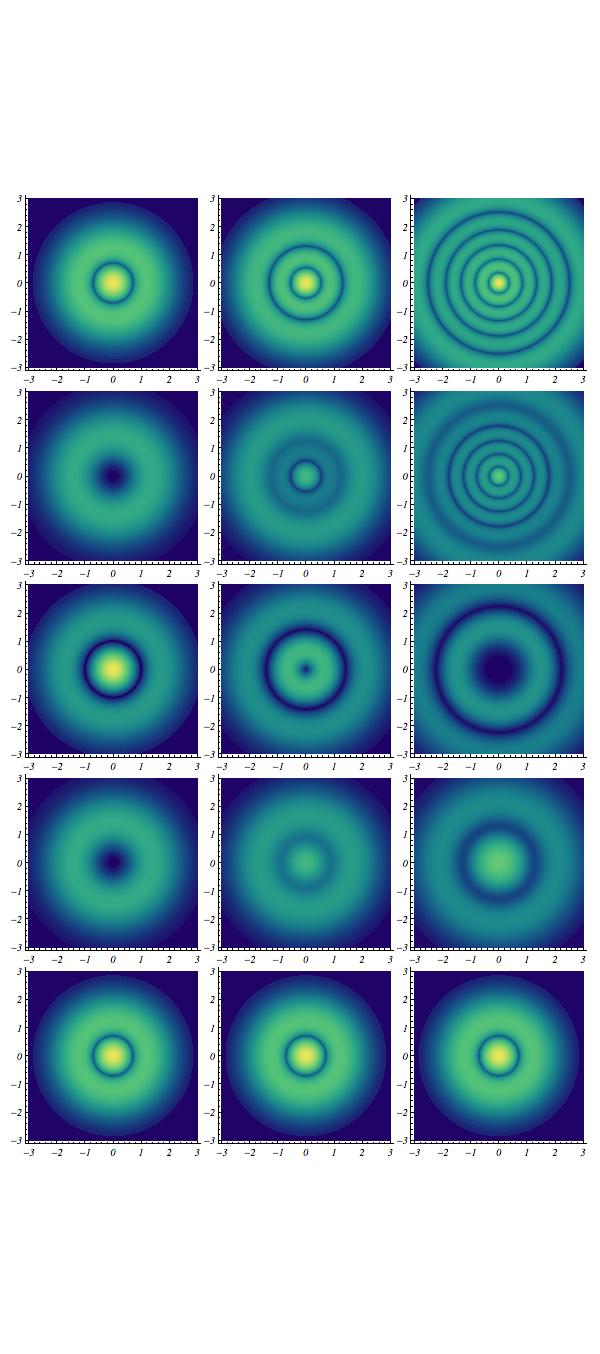}
\vspace{-2.2 cm}
\caption{\small  (Color online) Quantum beating for the NC harmonic oscillator described by the {\em traced-out} state vector $\tilde{\rho}^{(1)}_{n_{\tiny{x}},n_{\tiny{y}}}(\mathit{Q}_1,\Pi_1;t)$ in the $\mathit{Q}_1-\Pi_1$ plane, with $n_y = 2$ and $n_x = 1$ (first column), $2$ (second column) and $5$ (third column).
At time $\tau = 0$ the Wigner function is assumed to be centered at the origin $(0,\,0)$.
One has considered time intervals such that $\tau = k \pi(8\epsilon \Omega)^{-1}$, with $k = 0,\,1,\,2,\,3$ and $4$ in order to reproduce its time evolution through the sequence of five plots.
Notice that the prepared quantum state in $\mathit{Q}_1 - \Pi_1$ is projected into the {\em traced-out} quantum state, by reproducing its wave pattern at $\tau =  \pi/(2\epsilon \Omega)$.
It corresponds to the quantum beating effect with frequency $\omega_{beat} = 2\gamma = 2 \epsilon \Omega$.
The contour plot follows a {\em BlueGreenYellow} scale (from yellow (light gray) which corresponds to $1$, to blue (dark gray) which corresponds to $0$).}
\label{Mapa01}
\end{figure}

\begin{figure}
\vspace{-1.2 cm}
\includegraphics[width= 10cm]{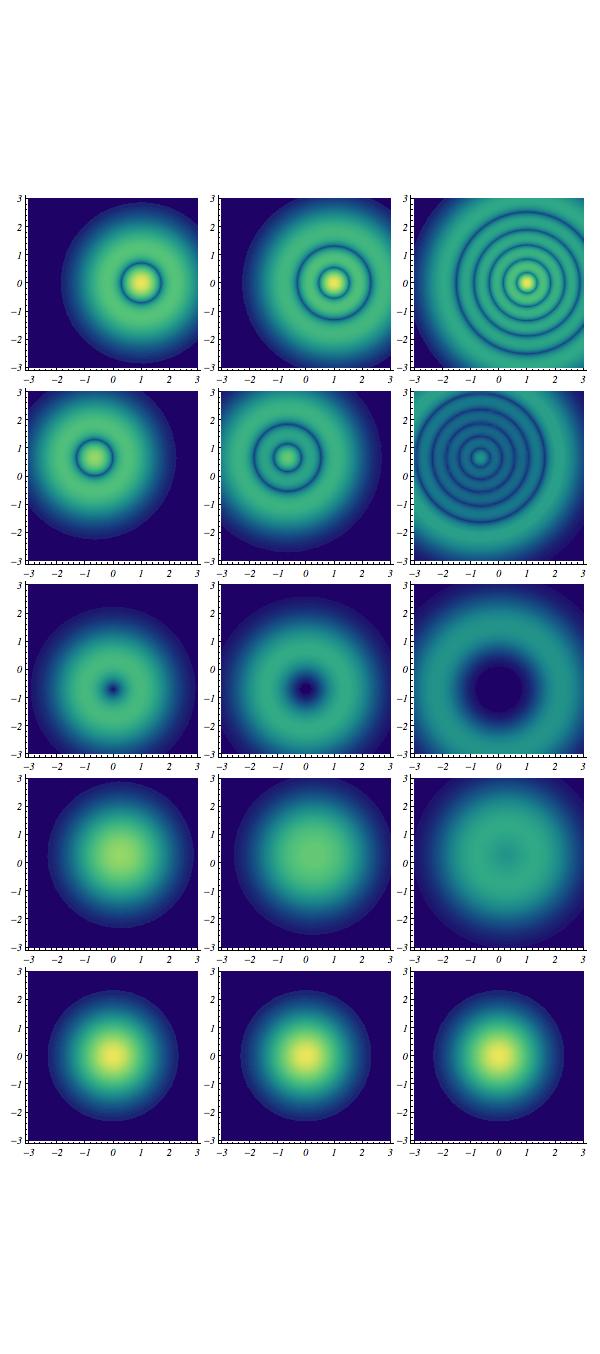}
\vspace{-2.2 cm}
\caption{\small (Color online) Quantum beating for the NC harmonic oscillator described by the {\em traced-out} state vector $\tilde{\rho}^{(1)}_{n_{\tiny{x}},n_{\tiny{y}}}(\mathit{Q}_1,\Pi_1;t)$ in the $\mathit{Q}_1-\Pi_1$ plane, with $n_y = 1$ and $n_x = 1$ (first column), $2$ (second column) and $5$ (third column).
The resulting Wigner function in the $\mathit{Q}_1 - \Pi_1$ plane  is supposed to be centered at $(\mathit{Q}_1,\Pi_1) = (1,\,0)$ at time $\tau = 0$.
One has considered time intervals such that $\tau = k \pi(8\epsilon \Omega)^{-1}$, with $k = 0,\,1,\,2,\,3$ and $4$.
Once again one notices a quantum beating effect with frequency $\omega_{beat} = 2\gamma = 2 \epsilon \Omega$.
Beside the analogous effects described in Fig.~\ref{Mapa01}, the NC parameter $\epsilon$ introduces the {\em precession} motion that follows the phase-space maps like those ones in Fig.~\ref{Phase}.}
\label{Mapa02}
\end{figure}

To show how the noncommutativity affects the state vectors prepared as pure states described by $\rho^W_{n_{\tiny{x}},n_{\tiny{y}}}$, one looks over the time evolution of the {\em traced-out} Wigner functions (density matrices) in the corresponding phase-space.
The {\em traced-out} Wigner functions are defined as
\begin{eqnarray}
\tilde{\rho}^{(1)}_{n_{\tiny{x}},n_{\tiny{y}}}(\mathit{Q}_1,\Pi_1;t) &=&  Tr_{\{2\}}\left[\rho^W_{n_{\tiny{x}},n_{\tiny{y}}}(\mbox{\bf \em Q},\mathbf{\Pi};t)\right]\nonumber\\
 &=& \int^{^{+\infty}}_{_{-\infty}}\hspace{-.5cm}d\mathit{Q}_2\int^{^{+\infty}}_{_{-\infty}}\hspace{-.5cm} d\Pi_2\,\, \rho^W_{n_{\tiny{x}},n_{\tiny{y}}}(\mbox{\bf \em Q},\mathbf{\Pi};t),\\
\tilde{\rho}^{(2)}_{n_{\tiny{x}},n_{\tiny{y}}}(\mathit{Q}_2,\Pi_2;t) &=&  Tr_{\{1\}}\left[\rho^W_{n_{\tiny{x}},n_{\tiny{y}}}(\mbox{\bf \em Q},\mathbf{\Pi};t)\right]\nonumber\\
 &=& \int^{^{+\infty}}_{_{-\infty}}\hspace{-.5cm}d\mathit{Q}_1\int^{^{+\infty}}_{_{-\infty}}\hspace{-.5cm} d\Pi_1\,\, \rho^W_{n_{\tiny{x}},n_{\tiny{y}}}(\mbox{\bf \em Q},\mathbf{\Pi};t).
\label{help05DD}
\end{eqnarray}
i. e. {\em tracing out} over $\mathit{Q}_{2,1}$ and $\Pi_{2,1}$ means integrating $\rho^W_{n_x,n_y}$ over these variables, so that the resulting Wigner function in the $\mathit{Q}_{1,2} - \Pi_{1,2}$ plane is obtained.
Assuming that we look over the time evolution of the state vectors along the $x$ direction, we consider the corresponding Wigner functions, $\tilde{\rho}^{(1)}_{n_{\tiny{x}},n_{\tiny{y}}}(\mathit{Q}_1,\Pi_1;t)$ as shown in Figs.~\ref{Mapa01} and \ref{Mapa02}.
At time $\tau = 0$, the Wigner function corresponds to a description of the 1D harmonic oscillator stationary state with quantum number $n_x$.
The quantum beating effect due to the NC parameter $\gamma$ is obtained by looking over time intervals multiples of $\pi(8\epsilon \Omega)^{-1}$.
The NC parameter $\epsilon = \gamma/\Omega$ modifies the usual time behavior of the commutative harmonic oscillator by generating the quantum beat with frequency $\omega_{beat} = 2 \gamma = 2\epsilon \Omega$.
Notice that the initial quantum state in the $\mathit{Q}_1 - \Pi_1$ phase subspace {\em collapses} into the {\em traced-out} quantum state, by reproducing its wave pattern at $\tau =  \pi/(2\epsilon \omega)$.
The results depicted in Figs.~\ref{Mapa01} and \ref{Mapa02} do not depend quantitatively on the auxiliary parameter, $\epsilon$, since we have chosen scale independent values for $\tau$.

To quantify the effect of the NC parameter, $\gamma$, over the time evolution of the state vectors like those from Eq.~(\ref{help05D}), one computes the linear entropy defined by 
\begin{eqnarray}
S_1(t) &=& 1 - \frac{2\pi}{\hbar} Tr_{\{1\}}\left[\left(Tr_{\{2\}}\left[\rho^W_{n_{\tiny{x}},n_{\tiny{y}}}(\mbox{\bf \em Q},\mathbf{\Pi};t)\right]\right)^2\right] \nonumber\\
&=& 1 - \frac{2\pi}{\hbar} Tr_{\{1\}}\left[\left(\tilde{\rho}^{(1)}_{n_{\tiny{x}},n_{\tiny{y}}}(\mathit{Q}_1,\Pi_1;t)\right)^2\right],\nonumber\\
S_2(t) 
&=& 1 - \frac{2\pi}{\hbar} Tr_{\{2\}}\left[\left(Tr_{\{1\}}\left[\rho^W_{n_{\tiny{x}},n_{\tiny{y}}}(\mbox{\bf \em Q},\mathbf{\Pi};t)\right]\right)^2\right] \nonumber\\
&=& 1 - \frac{2\pi}{\hbar} Tr_{\{2\}}\left[\left(\tilde{\rho}^{(2)}_{n_{\tiny{x}},n_{\tiny{y}}}(\mathit{Q}_2,\Pi_2;t)\right)^2\right], \nonumber\\
S_{12}(t) &=& 1 - \frac{4\pi^2}{\hbar^2} Tr_{\{1\}}\left[Tr_{\{2\}}\left[\left(\rho^W_{n_{\tiny{x}},n_{\tiny{y}}}(\mbox{\bf \em Q},\mathbf{\Pi};t)\right)^2\right]\right],
\label{linear}
\end{eqnarray}
through which the quantum {\em mutual information} is defined by
\begin{equation}
I_{12}(t) = S_1(t) + S_2(t) - S_{12}(t) = I_{21}(t),
\end{equation}
which quantifies the correlation between $x(\leftrightarrow 1)$ and $y(\leftrightarrow 2)$ states.
The mutual information is a measure of the correlation between subsystems of a quantum state.
If two variables $x$ and $y$ are assumed to be uncorrelated, the quantum {\em mutual information} measures the discrepancy in the uncertainty resulting from this possibly erroneous assumption.
By setting $\gamma = 0$, the mutual information, $I_{12}(t)$, as well as all the above defined entropies vanish, once $\rho^W_{n_{\tiny{x}},n_{\tiny{y}}}(\mbox{\bf \em Q},\mathbf{\Pi};t)$  reproduces the product of two uncorrelated pure states.
If the NC effects vanish, the state vector evolves like a stationary pure state.

Figs.~\ref{Mapa01} and \ref{Mapa02} also exhibit a swapping of states due to NC effects. 
It means that if one had prepared a 2D separable NC quantum state as $\rho^{W(t=0)}_{n_x,n_y} = \rho^W_{n_x}(\mathit{Q}_1(0),\Pi_1(0);0)\,\rho^W_{n_y}(\mathit{Q}_2(0),\Pi_2(0);0)$, which evolves in time, after a time $t = \tau = k \pi(8\epsilon \Omega)^{-1}$ (where $k$ is an integer number related to the beat frequency, $\omega_{beat} = 2\gamma = 2 \epsilon \Omega$) one should observe that the quantum state $\rho^{W(t=\tau)}_{n_x,n_y} = \rho^W_{n_x}(\mathit{Q}_1(\tau),\Pi_1(\tau);\tau)\,\rho^W_{n_y}(\mathit{Q}_2(\tau),\Pi_2(\tau);\tau)$ is converted into $\rho^{W(t=0)}_{n_y,n_x} = \rho^W_{n_y}(\mathit{Q}_1(0),\Pi_1(0);0)\,\rho^W_{n_x}(\mathit{Q}_2(0),\Pi_2(0);0)$, which corresponds to a swap of the quantum numbers, $n_x \leftrightarrow n_y$.
Fig.~\ref{Mapa01} illustrates such quantum number swapping in the $\mathit{Q}_1 - \Pi_1$ plane for $n_y = 2$ and $n_x = 1$ (first column), $2$ (second column) and $5$ (third column), 
and Fig.~\ref{Mapa02} does the same for $n_y = 1$ and $n_x = 1$ (first column), $2$ (second column) and $5$ (third column).

\begin{figure}
\subfigure[]{\includegraphics[width= 12cm]{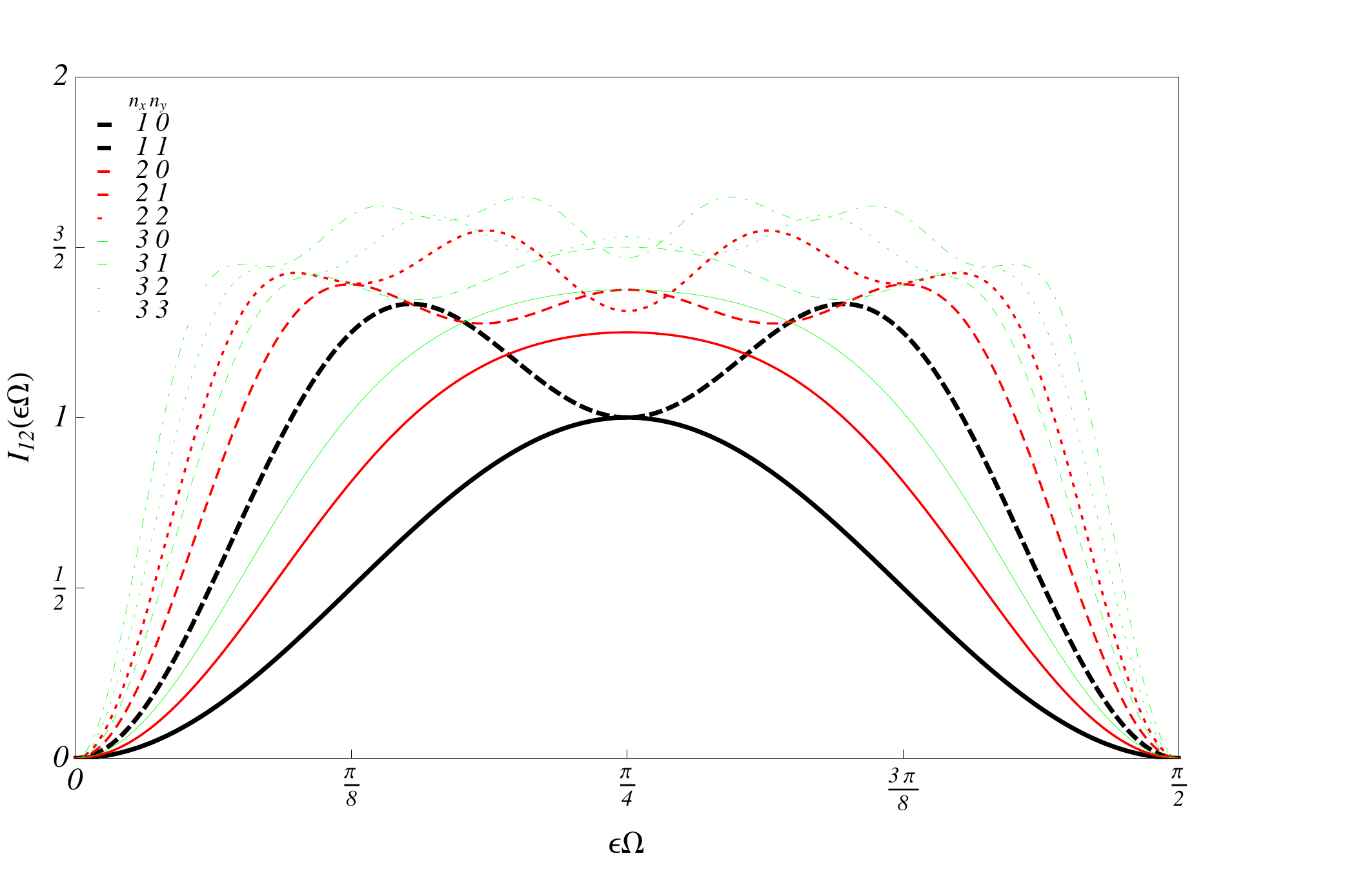}}
\subfigure[]{\includegraphics[width= 12cm]{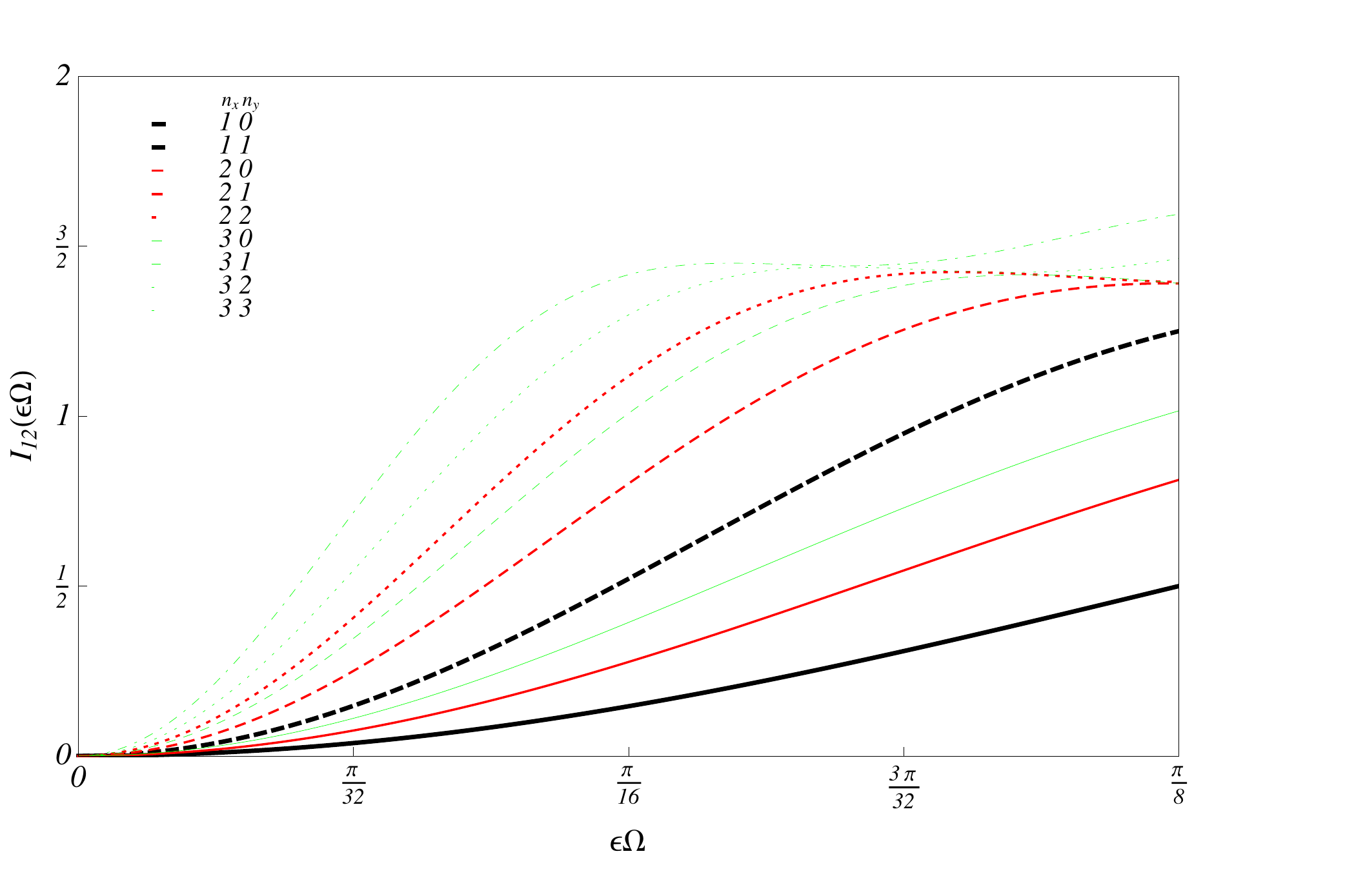}}
\caption{\small (Color online) (a) Mutual information, $I_{12}$, for the NC harmonic oscillator described by the state vector $\rho^W_{n_x,n_y}$ for several values of the quantum numbers $n_x$ and $n_y$. 
(b) Approximated decoherence profile for tiny values of the NC parameter $\epsilon$, depicted for increasing values of the mutual information, $I_{12}$.}
\label{Mutual02}
\end{figure}

It all works as if one had selected a specific {\em magnetic field}-like coupling between oscillators on the NC plane, which would exhibit the above-mentioned swapping of $x-y$ states.
The mutual information, $I_{12}$, as depicted from the first plot of Fig.~\ref{Mutual02}, quantifies the mutual interference between $x$ and $y$ states, which exclusively depends on the NC features.
This swapping dynamics is analogous to the one observed, for instance, in a weakly coupled QED cavity \cite{QEDcavity}.

In addition, if a time scale, $\tau$, is not sufficiently large to close the phase-space orbit (c. f. Fig.~(\ref{Phase})) for $\mathit{Q}_i(\tau)$ and $\Pi_i(\tau)$, the quantum beating can be reinterpreted as a scale-dependent decoherence effect as qualitatively illustrated by the second plot of Fig.~\ref{Mutual02}.
Depending on the magnitude of the NC parameter, $\gamma$, the beating frequency gives rise to a decoherence time, $\tau_{coh} = \pi /\gamma$.

The relevance of these results in terms of its experimental feasibility/detectability may depend essentially on the parameter $\eta$.
By setting $\eta = m^2\omega^2\theta$ one has $\hbar\gamma = \eta/m$ and the coherence time reduces to 
$\tau_{coh} = m h /2\eta$, % $\tau_{coh} = m [c^2]h /2\eta$,
such that only for small values of the mass, $m$, NC can play a relevant role in evidencing decoherence / quantum beating effects.
The corresponding coherence length related to $\tau_{coh}$ can be defined as
\begin{equation}
\lambda_{coh} = \frac{p}{m} \frac{1}{f} = \frac{\hbar}{\lambda m} \frac{1}{2\gamma} =\frac{\hbar \omega}{2\pi m} \frac{\hbar m}{2\eta} = \frac{\hbar^2 \omega}{4 \pi\eta}, %= \frac{\hbar^2 \omega}{\eta} [c],
\end{equation}
where it has been assumed that $c = 1$.

\section{Missing information and the thermodynamic limit}

A further issue to consider refers to the effect of noncommutativity on the loss of quantum coherence, the ensued missing quantum information and its imprint on the thermodynamic limit of quantum states when quantities like internal energy, Boltzmann entropy and heat capacity are investigated.

When one looks over a state vector in the NC plane $x-y$, in comparison with solutions of the commutative problem, the missing information can be quantified through the quantum entropy of the entire system in the Hilbert space.
The most useful variable for this analysis is the linear entropy, Eqs.~(\ref{linear}), computed over a thermalized statistical mixture of $n_1 - n_2$ states:
\begin{equation}
\rho^W_{th}(\sigma; r_i,k_i) = \frac{1}{\pi^2\, \hbar^2\,N(\sigma)}\, \exp[-\sigma]\,\sum_{n_1,n_2 = 0}^\infty \exp{\left[-(\sigma_1 n_1 + \sigma_2 n_2)\right]}\,\rho^W_{n_1,n_2}(r_i,k_i)
\label{SM}
\end{equation}
with
\begin{equation}
\sigma_s = \sigma (1-(-1)^s \epsilon),\qquad s = 1,\,2,
\end{equation}
where $\rho^W_{n_1,n_2}$ are the normalized state vectors given by Eq.~(\ref{Lague01}), $\sigma = \hbar\Omega/(k_B T)$, and $N(\sigma)$ is the normalization factor that contains the information from the corresponding statistical weights. 

This variable corresponds to the state operator of the infinite multiple modes $n_1$ and $n_2$ of the 2D harmonic oscillator, in equilibrium with a thermal bath through an interaction Hamiltonian factorizable  as $\mathcal{H}_1 \otimes \mathcal{H}_2\otimes \mathcal{H}_{bath}$ in the Hilbert space representation.

One notices that the state Eq.~(\ref{SM}) is quite suitable for analytical manipulations.
By introducing the coordinate parametrization,
\begin{equation}
r_i = \sqrt{\frac{\alpha}{\hbar\beta}}\mathit{Q}_i \quad \mbox{and} \quad k_i = \sqrt{\frac{\beta}{\hbar\alpha}}\Pi_i,
\end{equation}
such that $dr_i\,dk_i = d\mathit{Q}_i\, d\Pi_i$, with $i = 1,\,2$, 
and re-defining
\begin{equation}
\xi^2 = \sum_{i=1}^2\left(r_i^2+k_i^2\right), \nonumber
\end{equation}
\begin{equation}
\mathcal{L}=
\sum_{i,j = 1}^2{\left(\,\epsilon_{ij}\,k_i\, r_j\right)},\nonumber
\end{equation}
one can rewrite the statistical mixture Eq.~(\ref{SM}) as
\begin{eqnarray}
\rho^W_{th}(\sigma;\xi^2,\mathcal{L})&=& \frac{1}{\pi^2\, \hbar^2\,e^{\sigma}\,N(\sigma)}\,\exp{\left[-\xi^2\right]}\,\prod_{s=1}^2\left\{\sum_{n_s=1}^{\infty}
\left(-e^{-\sigma_s}\right)^{n_s}\,L^0_{n_s}\left[\xi^2+ 2(-1)^s \mathcal{L}\right]\right\}.
\end{eqnarray}

To obtain an explicit expression for $N(\sigma)$, one uses the following property of the Laguerre polynomials,
\begin{equation}
\sum_{l = 0}^{\infty} L^{\nu}_{l}(z)\Delta^{l} = \frac{1}{(1-\Delta)^{\nu+1}} \exp{\left[\frac{\Delta}{\Delta - 1}z\right]}; \quad|\Delta| < 1;
\label{laguerrrrre}
\end{equation}
through which, for $e^{-\sigma_s} < 1$, with $\Delta = -e^{-\sigma_s}$, one obtains
\begin{eqnarray}
\sum_{n_s=1}^{\infty}
\left(-e^{-\sigma_s}\right)^{n_s}\,L^0_{n_s}\left[\xi^2+ 2(-1)^s \mathcal{L}\right]=\frac{1}{1+e^{-\sigma_s}} \exp{\left[\frac{\xi^2+ 2(-1)^s \mathcal{L}}{1+e^{\sigma_s}}\right]},
\label{laguerrrrre2}
\end{eqnarray}
which results into
\begin{eqnarray}
\rho^W_{th}(\sigma;\xi^2,\mathcal{L})&=& \frac{1}{2 \,\pi^2\, \hbar^2\,N(\sigma)[\cosh{(\sigma)} + \cosh{(\epsilon\sigma)}]}\exp{\left[-\frac{\xi^2\,\sinh{(\sigma)} + 2 \mathcal{L}\,\sinh{(\epsilon\sigma)}}{\cosh{(\sigma)} + \cosh{(\epsilon\sigma)}}\right]}.
\end{eqnarray}

The normalization integral is evaluated over the infinite four-parameter phase-space, $\{\{r_1,k_1\},\{r_2,k_2\}\}$, and hence one obtains
\begin{equation}
N(\sigma) = \frac{1}{2}\frac{1}{\cosh{(\sigma)} - \cosh{(\epsilon\sigma)}},
\label{norm}
\end{equation}
such that
\begin{eqnarray}
\rho^W_{th}(\sigma;\xi^2,\mathcal{L})&=& 
\frac{1}{\pi^2\, \hbar^2}
\frac{\cosh{(\sigma)} - \cosh{(\epsilon \sigma)}}{\cosh{(\sigma)} + \cosh{(\epsilon \sigma)}}
\exp{\left[-\frac{\xi^2\,\sinh{(\sigma)} + 2 \mathcal{L}\,\sinh{(\epsilon\sigma)}}
{\cosh{(\sigma)} + \cosh{(\epsilon\sigma)}}\right]}.
\end{eqnarray}

\begin{figure}
\vspace{-2.2 cm}
\includegraphics[width= 5.5cm]{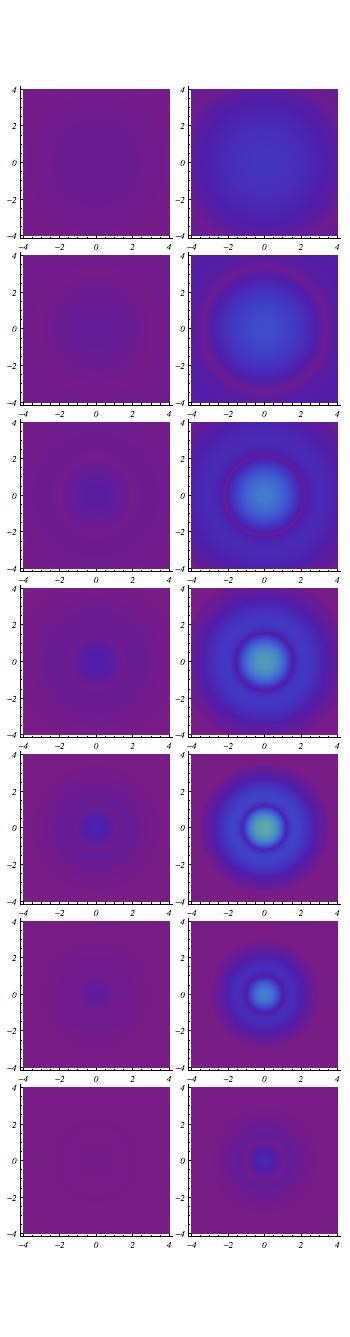}
\includegraphics[width= 5.5cm]{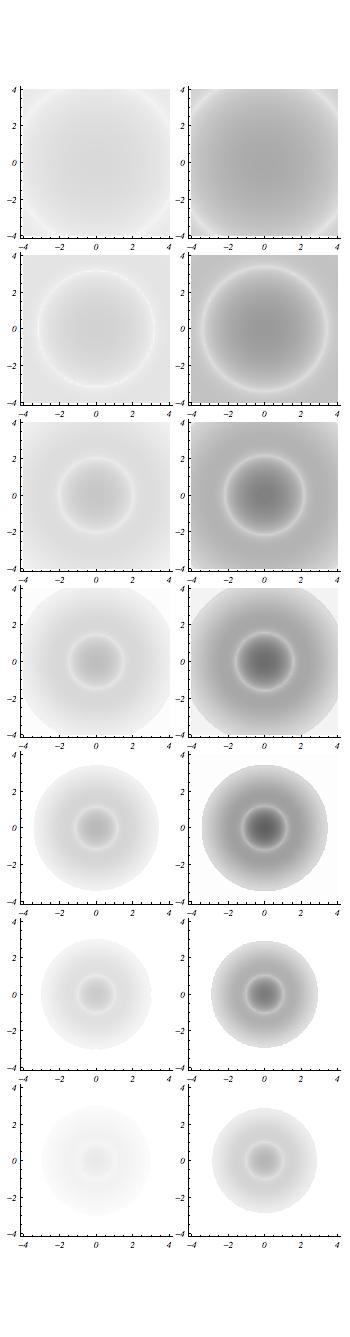}
\vspace{-1.4 cm}
\caption{\small (Color online) Probability distribution (quantum orbital) distortion for the NC harmonic oscillator state vectors with radial symmetry in the $\mathit{Q}_1-\mathit{Q}_2$ plane.
For higher temperatures, i. e. for $k_B T \gg \hbar\Omega$ ($\sigma \ll 1$), the NC effects are erased, and the quantum behavior of the thermalized state vector, $\rho^W_{th}$, prevails over the tiny NC quantum corrections. 
For very low temperatures, i. e. for $k_B T \ll \hbar\gamma$ ($\sigma \gg 1$), the wave function collapses into the classical limit.
The usual pattern of the quantum orbitals is maximally modified by the NC element at intermediate scales, namely $1 \lesssim  \sigma_{max} \lesssim 4$.
The plots correspond to $\epsilon  = 0.1$ ($1^{st}$-column) and $0.5$ ($2^{nd}$-column) with $\sigma$ assuming the values of $0.1,\,0.2,\,0.5,\,1,\,2,\,5$ and $10$ (decreasing temperature scale). 
The right-side {\em GrayLevel} plots correspond to the amplification of {\em monochrome negative image} (from white which corresponds to $0$ (no distortion), to black which corresponds to $1$ (maximal distortion)) of the distortion effects on the left-side plots.}
\label{Orbital}
\end{figure}

Since the above result depends on time-invariant quantities, $\xi^2$ and $\mathcal{L}$ (c. f. Eqs.~(\ref{invariant})), one can obtain the corresponding probability distribution relative to coordinates $\mathit{Q}_1$($\leftrightarrow r_1$) and $\mathit{Q}_2$($\leftrightarrow r_2$) by integrating the above thermalized Wigner function over $k_1$ and $k_2$, as
\begin{eqnarray}
\mathcal{P}(\sigma;r_1,r_2) = 
\int^{^{+\infty}}_{_{-\infty}}\hspace{-.5cm}d k_1
\int^{^{+\infty}}_{_{-\infty}}\hspace{-.5cm}d k_2\,\,
\rho^W_{th}(\sigma;\xi^2,\mathcal{L}). 
\end{eqnarray}
The distortion over the probability distribution resulting from NC effects can be obtained by subtracting $\mathcal{P}^{(\epsilon \rightarrow 0)}$ from $\mathcal{P}$.
The result follows from the axial invariance ($ \mathcal{P}(\sigma;r_1,r_2) \equiv \mathcal{P}(\sigma;Q^2) $) which can be visualized in the $\mathit{Q}_1-\mathit{Q}_2$ plane, as depicted from Fig.~\ref{Orbital}.
The results are described in terms of an increasing scale of the thermodynamic parameter $\sigma = \hbar\Omega/(k_B T)$.
For illustration purposes, the pictures in Fig.~\ref{Orbital} are exemplified for $\epsilon = \gamma/\Omega$ equals to $0.1$ and $0.5$.

\begin{figure}
\subfigure[Linear entropy and mutual information.]{\includegraphics[width= 12cm]{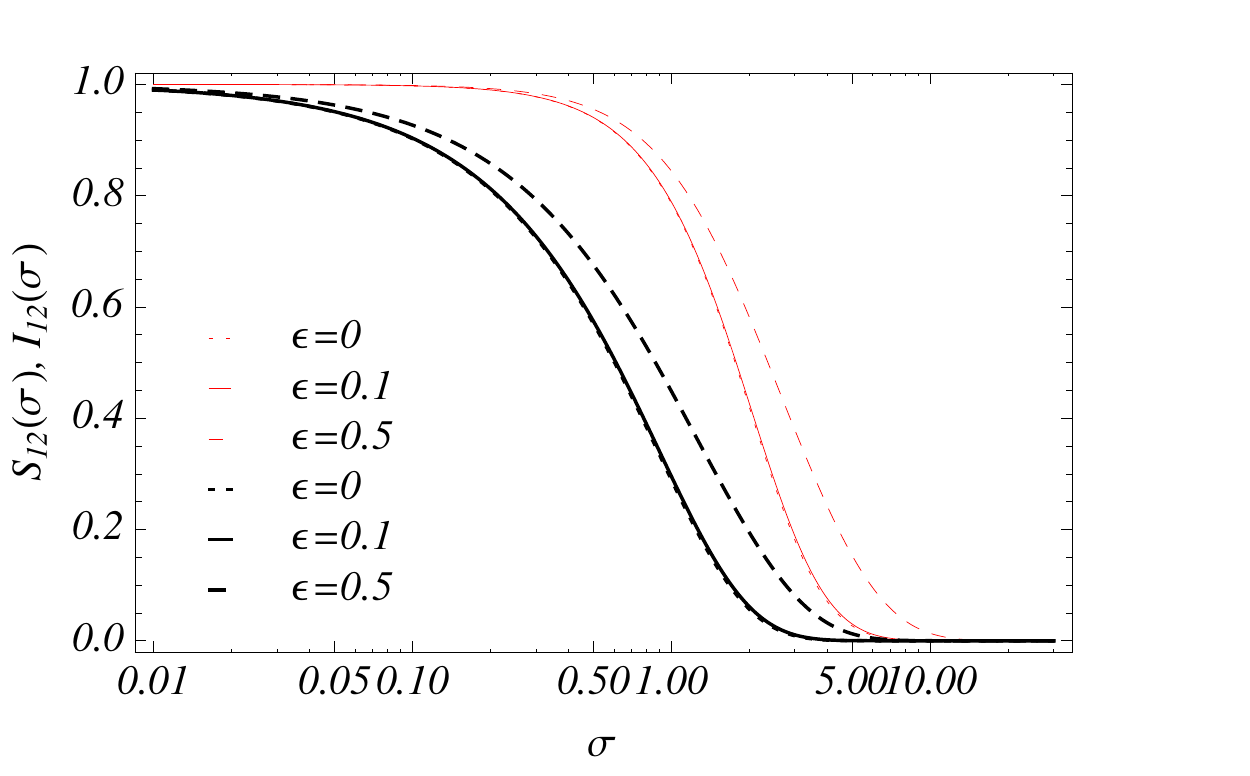}}
\subfigure[Missing information.]{\includegraphics[width= 12cm]{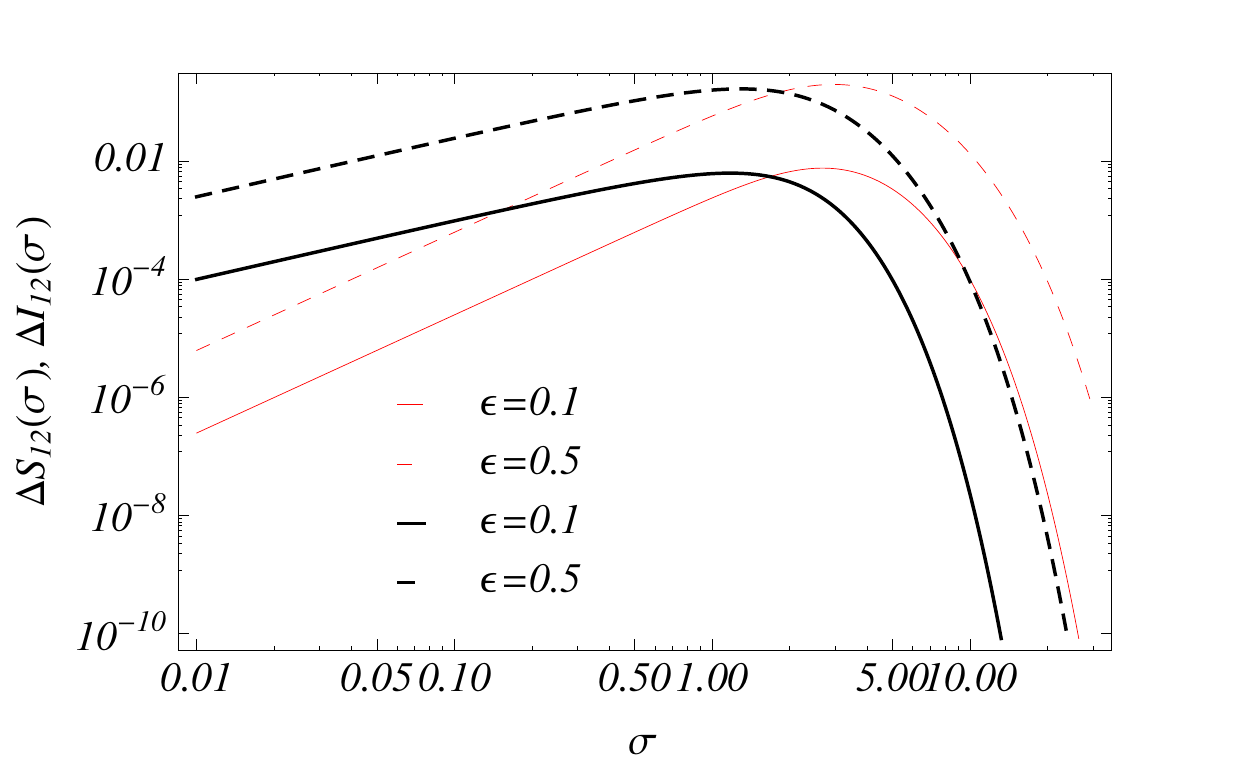}}
\caption{\small (Color online) (a) Linear entropy (thin red lines), $S_{12}(\sigma)$ and mutual information (thick black lines), $I_{12}(\sigma)$, for the isotropic 2D NC harmonic oscillator, and (b) the corresponding missing information given in terms of $\Delta S_{12}(\sigma)$ and $\Delta I_{12}(\sigma)$.
 At high temperatures, $k_B T \gg \hbar\gamma$ ($\sigma \ll 1$), the NC effects are erased.
At very low temperatures, $k_B T \ll \hbar\gamma$ ($\sigma \gg 1$), the wave function collapses into the classical limit.
The maximal values of $\Delta S_{12}(\sigma)$ and $\Delta I_{12}(\sigma)$ are obtained at intermediate values of $\sigma$ (see the correspondence in Fig.~\ref{Maximal}).
One has considered $\epsilon  = 0$ (dotted lines), $0.1$ (solid lines) and $0.5$ (dashed lines).}
\label{Entropy}
\end{figure}

To provide a more accurate estimate of the contribution of noncommutativity on the missing information, one examines the quantum entropy and the quantum mutual information of the system described above (c. f. the expression for $S_{12}$, Eq.~(\ref{linear})).
The linear entropy, $S_{12}(\sigma)$, and the mutual information, $I_{12}(\sigma)$, as well as the corresponding missing information described in terms of $\Delta S_{12}\sim S_{12}-S_{12}^{(\epsilon \rightarrow 0)}$ and $\Delta I_{12} \sim I_{12} - I_{12}^{(\epsilon \rightarrow 0)}$, are depicted in Fig.~\ref{Entropy}.

One notices that the isotropic 2D harmonic oscillator does exhibit an effect of noncommutativity through the statistical mixture, Eq.~(\ref{SM}). 

\begin{figure}
\includegraphics[width= 12cm]{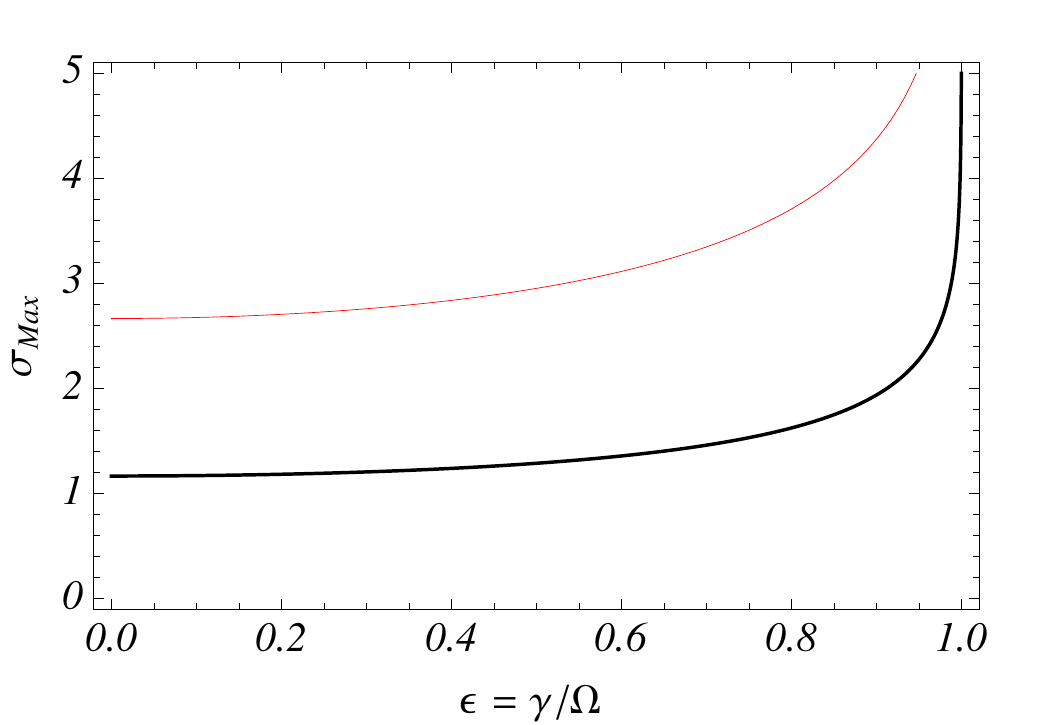}
\caption{\small (Color online) Numerical solution for the $\sigma$ parameter that maximizes the missing information (from Fig.~\ref{Entropy}) as function of the NC parameter $\epsilon = \gamma/\Omega$.
The results are for the maximal values of  $\Delta S_{12}(\sigma)$ (thin red line), and $\Delta I_{12}(\sigma)$ (thick black line).}
\label{Maximal}
\end{figure}

At high temperatures, the corrections due to noncommutativity reflect themselves into the missing information given by $\Delta S_{12} \propto \epsilon^2 \sigma^2$ and by $\Delta I_{12} \propto \epsilon^2 \sigma$, which are therefore small.
As the parameter $\sigma$ decreases, the system can access quantum states with larger quantum numbers, the linear entropy approaches the unity, and the system exhibits the behavior of a maximal statistical mixture, i. e. the classical limit.
For extremely low temperatures, $k_B T \ll \hbar\Omega$ ($\sigma \gg 1$), the system accesses predominantly quantum states with small quantum numbers.
This corresponds to a decreasing level of mixing and hence to quantum behavior, despite NC contributions.
For $\epsilon \ll 1$ ($\gamma \ll \Omega$) the usual pattern of missing information is maximally modified by the NC element at intermediate scales ($1 \lesssim  \sigma_{max} \lesssim 4$), as one can notice from numerical results depicted in Fig.~\ref{Maximal}.
There is a range for which NC effects are maximized.  

Qualitatively, the orbital distortion depicted in Fig.~\ref{Orbital} reflects an overall increase of the entropy of the system.
The orbital distortion describes how the NC quantum state superpositons deviate from the  commutative (standard QM) axially symmetric states.
Such deviations are effectively quantified by $\Delta S_{12}$ and $\Delta I_{12}$ as function of $\sigma$, which depends on the NC parameter $\epsilon$.
One can also see from Fig.~\ref{Entropy} that, in the limit where $\sigma \ll 1$, it corresponds to a largely suppressed smooth global change in the classical limit and, in the limit where $\sigma \gg 1$, it corresponds to suitable local changes on the commutative standard QM harmonic oscillator ground state
\footnote{Notice also that the {\em GrayLevel} map corresponds to a strong amplification by a power factor of $4$: squared amplitudes like $A^{-4}$ are depicted as $A^{-1}$ at the \em {GrayLevel} scale, for $A$ in the range $[0,1]$.}.

\subsection{The thermodynamic limit}

A large system of 2D NC harmonic oscillators in thermal contact with an environment, at temperature  $T$, corresponds to a {\em canonical} ensemble. 
The microstates occupied by the system are described by $\rho^W_{n_1,n_2}$, where $E_{n_1,n_2}$ denotes the ({\em stargen})energy of the system in a given microstate.
These microstates can be regarded as a system of discrete quantum states.

The classical partition function is obtained when the trace is expressed in terms of the coherent state vector, $\rho^W_{n_1,n_2}$, and when quantum uncertainties are negligible.
Formally, one has
\begin{eqnarray}
Z(\sigma) &=& 
\int^{^{+\infty}}_{_{-\infty}}\hspace{-.5cm}d r_1
\int^{^{+\infty}}_{_{-\infty}}\hspace{-.5cm}d r_2
\int^{^{+\infty}}_{_{-\infty}}\hspace{-.5cm}d k_1
\int^{^{+\infty}}_{_{-\infty}}\hspace{-.5cm}d k_2\,\,
\exp\left[-\frac{H^W_{HO}}{k_B\,T}\right] \,\rho^W_{n_1,n_2} 
\nonumber\\
&\equiv& \sum_{n_1,n_2 = 0}^\infty \exp\left[-\frac{E_{n_1,n_2}}{k_B\,T}\right] 
=\exp[{-\sigma}]\sum_{n_1,n_2 = 0}^\infty \exp\left[-\sigma(1 + \epsilon)n_1 - \sigma(1 - \epsilon)n_2\right]
\nonumber\\&=&  
\frac{e^{-\sigma}}{1 - e^{-\sigma(1 + \epsilon)}}\frac{1}{1 - e^{-\sigma(1 - \epsilon)}} \nonumber\\
&=& \frac{1}{2}\frac{1}{\cosh{(\sigma)} - \cosh{(\epsilon\sigma)}}
\quad = \quad N(\sigma),
\label{partition}
\end{eqnarray}
which is consistent with Eq.~(\ref{norm}).

The partition function, $Z(\sigma)$, allows for obtaining the thermodynamic variables of the system, such as the internal energy:
\begin{equation}
U(\sigma) = - \hbar \Omega \,\frac{\partial}{\partial \sigma} \ln{\left[Z(\sigma)\right]};
\end{equation}
the Boltzmann entropy:
\begin{equation}
S_k(\sigma) = -k_B \,\left\{\ln{\left[Z(\sigma)\right]} - \sigma U(\sigma)/\hbar \Omega\right\};
\end{equation}
and the heat capacity:
\begin{equation}
C_v = k_B \,\sigma^2\,\frac{\partial^2}{\partial \sigma^2}\ln{\left[Z(\sigma)\right]}.
\end{equation}

By substituting the result from Eq.~(\ref{partition}) into the above definitions, one obtains
\begin{equation}
U(\sigma) = - \hbar \Omega  \frac{\sinh{(\sigma)} - \epsilon \sinh{(\epsilon\sigma)}}
{\cosh{(\sigma)} - \cosh{(\epsilon\sigma)}},
\end{equation}
\begin{equation}
S_k(\sigma) = - k_B 
\frac{\left(\cosh{(\sigma)} - \cosh{(\epsilon\sigma)}\right)
\,\ln\left[2\cosh{(\sigma)} - 2\cosh{(\epsilon\sigma)}\right]
+ \sigma \left[ \sinh{(\sigma)} - \sinh{(\epsilon\sigma)}\right]}
{\cosh{(\sigma)} - \cosh{(\epsilon\sigma)}},
\end{equation}
and
\begin{equation}
C_v = k_B\frac{\sigma^2\,(1 +\epsilon^2)\left[\cosh{(\sigma)}\cosh{(\epsilon\sigma)} - 1 \right]
- 2 \sigma^2\, \epsilon \sinh{(\sigma)}\sinh{(\epsilon\sigma)}}{\left(\cosh{(\sigma)} - \cosh{(\epsilon\sigma)}\right)^2},
\end{equation}
from which the NC corrections are displayed through the explicit dependence on $\epsilon = \gamma /\Omega$.

\begin{figure}
\subfigure[Thermodynamic variables.]{\includegraphics[width= 12cm]{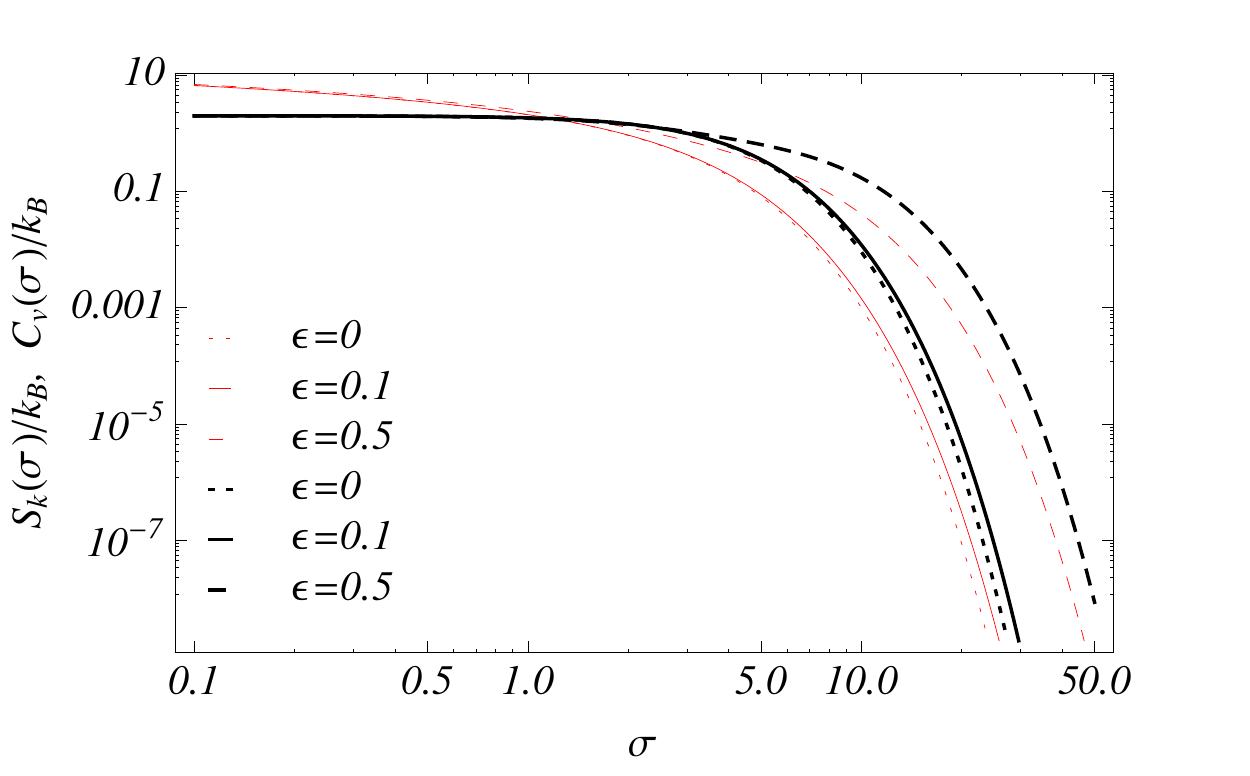}}
\subfigure[Distortions due to NC effects.]{\includegraphics[width= 12cm]{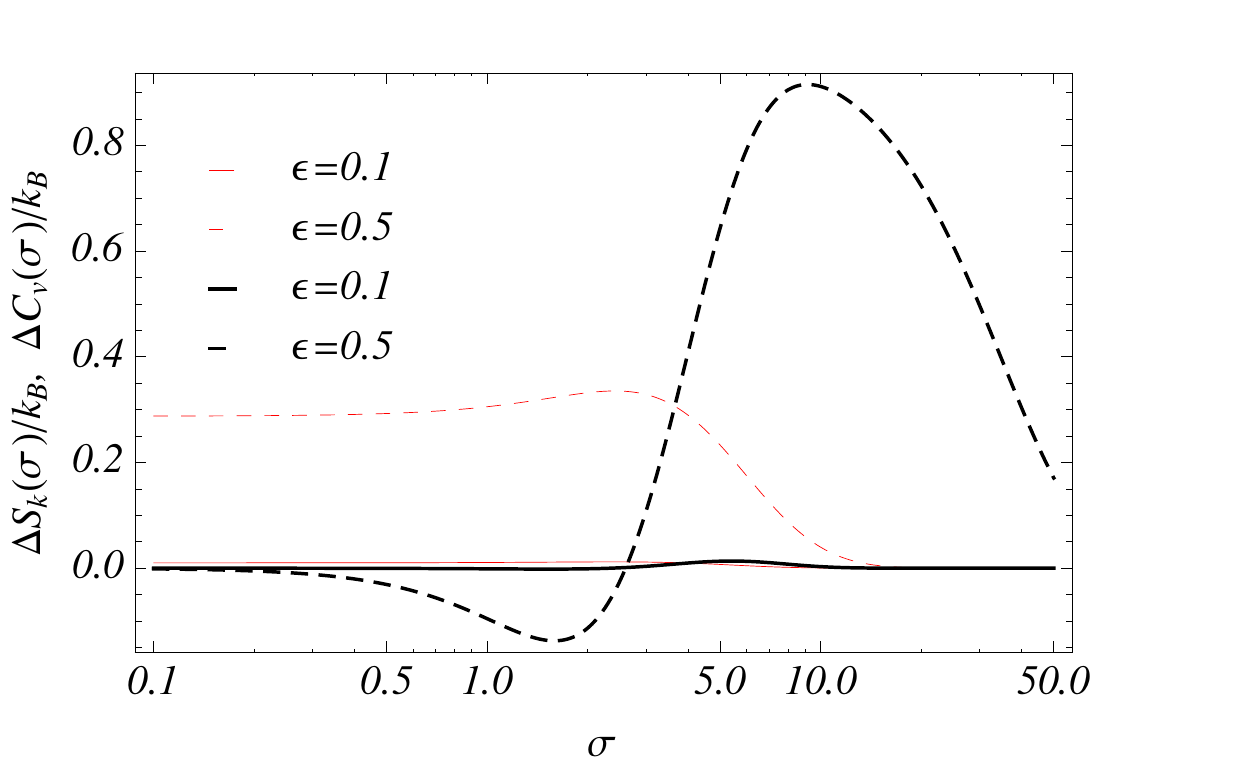}}
\caption{\small (Color online) (a) Boltzmann entropy (thin red lines), $S_k(\sigma)/k_B$, and heat capacity (thick black lines), $C_v(\sigma)/k_B$, for the statistical mixture $\rho^W_{th}$.
It reproduces the thermodynamic limit of the isotropic 2D NC harmonic oscillator quantum behavior.
The distortion due to the NC effects are quantified through $\Delta S_k(\sigma)/k_B$ and $\Delta C_v(\sigma)/k_B$ from plot (b).
One notices that $C_v(\sigma)/k_B$ exhibits an anomalous behavior that inverts the sign of $\Delta C_v(\sigma)/k_B$ at intermediate scales of $\sigma$.
The plots are for $\epsilon  = 0$ (dotted lines), $0.1$ (solid lines) and $0.5$ (dashed lines).}
\label{Heat}
\end{figure}

The NC effects on the above defined thermodynamic variables, $S_k$ and $C_v$, are depicted in Fig.~\ref{Heat}.
The results for the Boltzmann entropy, $S_k$, shows that the NC features increase the entropy.
The heat capacity, $C_v$, shows an anomalous dependence on the thermodynamic parameter $\sigma$ such that the NC corrections may increase or decrease with respect to $C_v^{(\epsilon \rightarrow 0)}$ (c. f. the second plot of Fig.~\ref{Heat}).
For these thermodynamic variables, the effect of noncommutativity is maximal for intermediate scales of $\sigma$.
This offers a novel possibility for measuring NC effects at low temperatures.

\section{Axial symmetry and the $\eta$-Zeeman effect}

It can be speculated whether the NC effects exhibited by the isotropic harmonic oscillator can be factorized from the Wigner distribution if one chooses the appropriate representation for the quantum operators.
Presumably, the axial symmetry of the problem might be relevant to obtain a suitable representation.

From the point of view of the quantum system described by the Hamiltonian, Eq.~(\ref{Hamilton}), the noncommutativity can be re-interpreted as an analog of the Zeeman effect for a charged rotating particle in the presence of an external magnetic field $\mathbf{B}$.

In this context, it is helpful to consider the three-dimensional (3D) extension of the previously discussed NC problem.

The Hamiltonian for the problem can be written as
\begin{eqnarray}
&\hat{H}_{HO}(\mathbf{q},\mathbf{p}) &= \frac{\mathbf{p}^2}{2m} + \frac{1}{2}m \omega^2 \mathbf{q}^2,\nonumber\\
\Rightarrow &H^W_{HO}(\mbox{\bf \em Q},\mathbf{\Pi}) &= \frac{1}{2m}\left[\mathbf{\Pi} + \frac{1}{2\hbar}\left({\boldsymbol\eta}\times \mbox{\bf \em Q} \right)\right]^2 + \frac{m\omega^2}{2} 
\left[\mbox{\bf \em Q} - \frac{1}{2\hbar}\left({\boldsymbol\theta}\times \mathbf{\Pi} \right)\right]^2,
\label{Hamiltonian2A}
\end{eqnarray}
since we have considered a simplified version of the 3D SW map given by
\begin{equation}
 \hat{q}_i =  \mathit{\hat{Q}}_i - {1\over 2 \hbar} \epsilon_{ijk} \theta_j{\hat{\Pi}}_k \hspace{0.5 cm},\hspace{0.5 cm}  \hat{p}_i =  \hat{\Pi}_i + {1\over 2 \hbar} \epsilon_{ijk}  \eta_j\,\mathit{\hat{Q}}_k~,
\label{SWmap2}
\end{equation}
that reflects the following NC rules,
\begin{equation}
\left[  \hat{q}_i,  \hat{q}_j \right] = i \epsilon_{ijk}\theta_{k} \hspace{0.2 cm}, \hspace{0.2 cm} \left[  \hat{q}_i,  \hat{p}_j \right] = i \delta_{ij}\hbar \left(1 - \frac{{\boldsymbol\theta}\cdot{\boldsymbol\eta}}{4\hbar^2}\right)\hspace{0.2 cm}, \hspace{0.2 cm} \left[  \hat{p}_i,  \hat{p}_j \right] = i \epsilon_{ijk}\eta_{k}\hspace{0.2 cm}, \hspace{0.2 cm} i,j=1,2,3
\end{equation}
where $\epsilon_{ijk}$ is the Levi-Civita tensor of rank three, ${\boldsymbol \theta} = (\theta_1,\theta_2,\theta_3)$, and ${\boldsymbol\eta} = (\eta_1,\eta_2,\eta_3)$.
The 2D version of the problem can be recovered from above relations by setting $\theta_{1,2}=\eta_{1,2} = 0$, $\theta_3 \equiv \theta$, and $\eta_3 \equiv \eta$.  

Even for a free particle, NC effects on the Weyl-Wigner Hamiltonian, $H^W_{HO}$, will be present in the particle dynamics through its kinetic energy:
\begin{eqnarray}
&\hat{H}_{HO}(\mathbf{p}) &= \frac{\mathbf{p}^2}{2m},\nonumber\\
\Rightarrow &H^W_{\eta}(\mbox{\bf \em Q},\mathbf{\Pi}) & = \frac{1}{2m}\left[\mathbf{\Pi} + \frac{1}{2\hbar}\left({\boldsymbol\eta}\times \mbox{\bf \em Q} \right)\right]^2\nonumber\\
&&=  \frac{\mathbf{\Pi}^2}{2m} + 
\frac{1}{2m\hbar}{\boldsymbol\eta}\cdot \left(\mbox{\bf \em Q} \times \mathbf{\Pi}\right) + \frac{1}{8m\hbar^2}\left({\boldsymbol\eta} \times \mbox{\bf \em Q}\right)^2.
\label{Hamiltonian2B}
\end{eqnarray}

One can use that
\begin{equation}
{\boldsymbol\eta}\cdot \left(\mbox{\bf \em Q} \times \mathbf{\Pi}\right) = {\boldsymbol\eta}\cdot \mathbf{L} = \eta L \cos{(\varphi)} = \eta L_{z},
\end{equation}
and
\begin{equation}
\left({\boldsymbol\eta} \times \mbox{\bf \em Q}\right)^2 = \eta^2 R^2,
\end{equation}
with $R = |\mbox{\bf \em Q}| \sin{(\varphi)} =  \sqrt{\mathit{Q}_1^2 + \mathit{Q}_2^2}$, where $\varphi = \arctan{R/\mathit{Q}_3}$ is the angle measured on the plane of axial symmetry along $\hat{z}$, defined by an arbitrary direction of ${\boldsymbol \eta}$ (that in this case corresponds to ${\boldsymbol \eta} = \eta \hat{z}$).
The axial symmetry ensures that the Hamiltonian Eq.~(\ref{Hamiltonian2B}) can be rewritten as
\begin{eqnarray}
H^W_{\eta}(\mathbf{\Pi},\mbox{\bf \em Q}) &=&  \frac{\mathbf{\Pi}^2}{2m} + 
\frac{\eta}{2m\hbar} L_{z} + \frac{\eta^2}{8m\hbar^2} R^2.
\label{Hamiltonian2C}
\end{eqnarray}

Dropping the $\mathbf{q}$-dependent potential, the Moyal {\em star}-product properties lead to the {\em star}genvalue equation written as
\begin{eqnarray}
\hat{H}_{HO}(\mathbf{p}) \star \Psi(\mathbf{q}) 
%\equiv H_{\eta}(\mathbf{p}) \star_{\eta} \star_{\hbar} \Psi(\mathbf{q}) 
=  H^W_{\eta}(\mathbf{\Pi},\mbox{\bf \em Q}) \star \Psi(\mbox{\bf \em Q}) = E\, \Psi(\mbox{\bf \em Q}).
\label{Hamiltonian3}
\end{eqnarray}
The second-order correction due to the $\eta^2$-term has to be considered in order to set a complete map of the {\em Zeeman} effect \cite{Ballentine}.
The replacement of the parameter $\eta$ by the coefficient $e B \hbar /c$ into Eq.~(\ref{Hamiltonian2C}) maps into the dynamics of a particle with charge, $e$, moving in a plane perpendicular to the magnetic field $\mathbf{B} = B \hat{z}$.

The eigenfunctions are obtained in terms of cylindrical coordinates ($R,\varphi,z$) as $\Psi(\mbox{\bf \em Q}) = \psi_{\kappa,\ell}(R,\varphi)\,\mathcal{Z}(z)$.
Since $\mathcal{Z}(z) \propto \exp{(-i k_z z/\hbar)}$ is not relevant for the present purpose, one has
\begin{eqnarray}
 H^W_{\eta}(\mathbf{\Pi},\mbox{\bf \em Q}) \star \Psi(\mbox{\bf \em Q}) &=&  \left(\tilde{H}^W_{\eta}(R,\varphi) - \frac{\hbar^2}{2m}\frac{\partial^2}{\partial z^2}\right) \psi_{\kappa,\ell}(R,\varphi)\,\mathcal{Z}
\nonumber\\ &=& \left(E_{\kappa,\ell} + \frac{k_z^2}{2m}\right)\psi_{\kappa,\ell}(R,\varphi)\,\mathcal{Z}.
\label{Hamiltonian4}
\end{eqnarray}

The remaining eigenvalue equation becomes
\small
\begin{eqnarray}
\tilde{H}^W_{\eta}(R,\varphi)  \psi_{\kappa,\ell}(R,\varphi) &=&
\left[
-\frac{\hbar^2}{2m}\left(\frac{\partial^2}{\partial R^2} +\frac{1}{R}\frac{\partial}{\partial R} + \frac{1}{R^2}\frac{\partial^2}{\partial \varphi^2}\right)
-i \frac{\eta}{2m}\frac{\partial}{\partial \varphi} + \frac{\eta^2}{8m\hbar^2} R^2 \right]\psi_{\kappa,\ell}(R,\varphi)
\nonumber\\
&=& E_{\kappa,\ell}\,  \psi_{\kappa,\ell}(R,\varphi),
\label{Hamiltonian5}
\end{eqnarray}
\normalsize
where we have identified $\Pi_i = -i \hbar (\partial/\partial \mathit{Q}_i)$, with $i = 1,2$, and $L_z = -i \hbar (\partial/\partial \varphi)$, and the simultaneous eigenfunction of $\tilde{H}^W_{\eta}$ and $L_z$, $\psi_{\kappa,\ell}(R,\varphi)$, can be obtained by separation of variables as $\psi_{\kappa,\ell}(R,\varphi) = f_{\kappa,\ell}(R) \,e^{i\,\ell \varphi}$.

The resulting ordinary differential equation for $f_{\kappa,\ell}(R)$ is given by
\begin{equation}
\left[
-\frac{\hbar^2}{2m}\left(\frac{\partial^2}{\partial R^2} +\frac{1}{R}\frac{\partial}{\partial R} - \frac{\ell^2}{R^2}\right)
+ \frac{\ell\eta}{2m} + \frac{\eta^2}{8m\hbar^2} R^2 - E_{\kappa, \ell}\right]f_{\kappa,\ell}(R) = 0,
\label{Hamiltonian6}
\end{equation}
for which, after some mathematical manipulations, we obtain the solution
\begin{equation}
f_{\kappa,\ell}(R) = \left(\frac{\eta}{2\hbar^2} R^2\right)^{\frac{|\ell|}{2}} \, 
\exp{\left(-\frac{\eta}{4\hbar^2}R^2\right)} \, L^{|\ell|}_{\kappa}\left(\frac{\eta}{2\hbar^2}R^2\right),
\label{Hamiltonian7}
\end{equation}
and the energy spectrum depending on the quantum numbers, $\kappa$ and $\ell$:
\begin{equation}
E_{\kappa,\ell} = \frac{\eta}{2m}\left(2\kappa + |\ell| - \ell + 1\right).
\end{equation}

If one considers the harmonic oscillator potential at the NC Hamiltonian from Eq.~(\ref{Hamiltonian2A}), the result of the above analysis would be valid up to {\em second-order} in $\theta$ and $\eta$, and would lead to the energy spectrum:
\begin{equation}
E_{\kappa,\ell} = 2\hbar\alpha\beta (2\kappa + |\ell| + 1) - \hbar \gamma \ell,
\end{equation}
which relates to the solution described in the previous sections through $2\kappa + |\ell| = n_{1} + n_{2}$ and  $\ell = n_{2} - n_{1}$.
The normalized eigenfunction, in that case, would be given by
\begin{equation}
\psi_{\kappa,\ell}(R,\varphi) = \left(\frac{\alpha}{\hbar\beta} R^2\right)^{\frac{|\ell|}{2}} \, 
\exp{\left(-\frac{\alpha}{\hbar\beta}R^2\right)} \, L^{|\ell|}_\kappa\left(\frac{\alpha}{\hbar\beta}R^2\right)\, \exp{(i\,\ell\, \varphi)}.
\label{Hamiltonian8}
\end{equation}
with $\alpha$ and $\beta$ redefined up to second-order in $\theta$ and $\eta$.
The relevance of separating the solution in cylindrical coordinates is that the re-defined quantum numbers introduce energy eigenfunctions and eigenvalues with well-stablished physical meaning.
To sum up, the analytic solution of the problem that contains second-order corrections in $B$ (or, equivalently, $\eta^2$, in the NC version) is provided in terms of associated Laguerre functions $L^{|\ell|}_\kappa$. 
Notice that the solution $f_{\kappa}(R)$ has been presented in terms of a regular power series expansion (see, for instance, the resolution of problem 11.6, pag. 632, at Ref.~\cite{Ballentine}), and it is also discussed in Ref.~\cite{Banerjee}.

\section{Conclusions}

In this work we have examined the implications of noncommutativity on quantum beating, scale dependent decoherence and dissipation effects related to the loss of information of quantum systems.
Considering the time-evolution of the 2D harmonic oscillator in the NC phase-space through the Groenwold-Moyal description of the Wigner function, we have reconstructed a suitable framework for investigating NC effects.
It is shown that the harmonic oscillator on the NC plane approaches the classical limit, and exhibits well marked quantum effects like state-swapping, quantum beating, and loss of quantum coherence to some extent.

We have shown that phase-space noncommutativity may induce the destruction of the original features of the wave function for the commutative problem.
The NC terms distort the 2D harmonic oscillator quantum orbitals through modifications onto their corresponding probability distributions.
To study these features, the missing information quantified by the linear entropy was computed in the NC context.
The linear entropy and the quantum mutual information of the 2D NC harmonic oscillator were examined and the conditions for identifying the maximal missing information due to the phase-space noncommutativity was obtained.

Finally, we have also investigated the thermodynamic limit of the quantum system by discussing the behavior of the Boltzmann entropy and of the heat capacity derived from the partition function obtained for the thermalized 2D NC harmonic oscillator.
A correspondence between the NC effects over quantum and classical variables were evidenced, and it was shown that the most suitable range of temperatures for detecting noncommutativity is around $\sim \hbar \Omega/ k_B$ (c. f. Eq.~(\ref{eq37})).

As a complementary issue, the 2D NC harmonic oscillator problem was rewritten in order to depict an axial symmetry so to ensure a suitable factorization of the NC effects.
An analogy with Zeeman-effect has been drawn, resulting in a description of the system with well-defined NC quantum number and a factorizable wave-function.
The results allow us to decouple the noncommutative effects such that it resembles a Zeeman-like effect.

One additional feature of the axial symmetry of the problem deserves some comments.
The introduction of an intrinsic anisotropy into the quadratic potential of the NC harmonic oscillator would not qualitatively change the obtained results.
Besides suppressing the possibility of decoupling the NC effects that break the isotropy through the introduction of different elastic constants, $k_1\propto\omega_1^2$ and $k_2\propto\omega_2^2$, would lead to different quantum beating patterns.
In this case, in order to recover the initial configuration of the phase-space, the system would require a number of cycles proportional to the minimal common integer multiple of oscillation frequencies $\Omega_1$, $\Omega_2$, $\gamma_1$ and $\gamma_2$.

%One could also consider the possibility of introducing a linear perturbation to the harmonic oscillator in the coordinate variables ($Q_1$ or $Q_2$) in order to produce some kind of anisotropy between $Q_1$ and $Q_2$.
%Actually, this would only produce a relative shift in the phase-spaces ($(Q_1,\Pi_2)$ or $(Q_2,\Pi_1)$) likewise the effect of gravity in the unidimensional harmonic oscillator in $\hat{z}$ direction.

To summarize, our results suggest that noncomutativity effects can be interestingly considered when addressing the issues of interface between quantum and classical descriptions of nature \cite{14to15}.
As it happens, for quantum superpositions, that are irreversibly affected by their surroundings, the quantum collapse due to noncommutativity may reflects the loss of coherence of pure states that are continuously transformed into a statistical mixture. 
In the general context of the discussion of mechanisms to induce decoherence and loss of information \cite{17to22}, noncommutativity is shown to play a relevant role in inducing the transition from quantum to classical behavior. 

\vspace{.5 cm}
{\em Acknowledgments - The work of AEB is supported by the Brazilian Agencies FAPESP (grant 12/03561-0) and CNPq (grant 300233/2010-8). The work of O. B. is partially supported under the Portuguese Funda\c{c}\~ao para Ci\^encia e Tecnologia (FCT) by the project PTDC/FIS/11132/2009}.

\pagebreak
\newpage

\renewcommand{\baselinestretch}{1}

\end{document}